\documentstyle[aps,prb,epsf,rotate,multicol,color,float]{revtex}

\definecolor{red}{rgb}{1,0,0}
\definecolor{blue}{rgb}{0,0,1}
\definecolor{green}{rgb}{0,1,0}
\newcommand{\red}{\textcolor{red}}
\newcommand{\blue}{\textcolor{blue}}
\newcommand{\green}{\textcolor{green}}

\newcommand{\veta}{\mbox{\boldmath$\eta$}}

\newcommand{\lrule}{ \noindent
        \rule{0.5\textwidth}{0.1mm}\rule{0.1mm}{3pt}\newline }

\begin{document}
\draft
\title{Antiferromagnetic Domains and
Superconductivity in UPt$_3$
}

\author{Matthias J. Graf}
\address{ Theoretical Division\\
Los Alamos National Laboratory,
Los Alamos, New Mexico \ 87545 
}
\author{and \\[2ex] Daryl W. Hess}
\address{ Center for Computational Materials Science\\
Naval Research Laboratory,
Washington, D.C. \ 20375-5345
}
\date{December 22, 2000}
%\date{\today}
\maketitle  

\begin{abstract}
We explore the response of an unconventional superconductor to spatially 
inhomogeneous antiferromagnetism (SIAFM). Symmetry allows the superconducting 
order parameter in the E-representation models for UPt$_3$ to couple directly 
to the AFM order parameter.  The Ginzburg-Landau equations for coupled 
superconductivity and SIAFM are solved numerically for two possible SIAFM 
configurations: (I) abutting antiferromagnetic domains of uniform size,
and (II) quenched random disorder of `nanodomains' in a uniform AFM 
background.  
We discuss the contributions to the free energy, specific heat, and order
parameter for these models.  Neither model provides a satisfactory account of 
experiment, but results from the two models differ significantly. 
Our results demonstrate that the response of an E$_{2u}$ superconductor
to SIAFM is strongly dependent on the spatial dependence of AFM order; 
no conclusion can be drawn regarding the compatibility of E$_{2u}$ 
superconductivity with UPt$_3$ that is independent of assumptions
on the spatial dependence of AFM.
\end{abstract}

\pacs{PACS numbers: 74.70.Tx, 74.20.-z, 74.80.-g, 74.25.Dw, 74.25.Ha
\hfill
LA-UR: 00-3510}

\begin{multicols}{2}

\section{Introduction}

  The nature of the spatially inhomogeneous small-moment antiferromagnetism 
 (AFM) observed in neutron scattering experiments\cite{neutrons,isaacs95,hayden}
 below $T_{N} \sim 6$K and its interaction 
 with superconductivity ($T_c \sim 0.5$K) remain central issues in determining 
 the symmetry of the superconducting order parameter of UPt$_3$.  The unusual
 H-T phase diagram\cite{adenwalla90,bruls90} at ambient pressure apparently 
 shows three superconducting phases in the mixed state and two Meissner phases. 
 Experimental studies using hydrostatic\cite{hayden} and uniaxial\cite{uniax} 
 pressure reveal the existence of a critical pressure above which the 
 zero-field transition splitting disappears.
 This complex phase diagram strongly suggests that superconductivity in 
 this heavy-electron material is unconventional and has provided motivation for
 much theoretical work.\cite{sauls94,heffner96,machida89,chen93,cox95,zhitomirsky96}  
 Proposed theories range from odd-in-frequency 
 pairing\cite{cox95} to multicomponent order parameters. 
 The latter may belong to a single multidimensional representation 
 of the symmetry group (see Refs.\onlinecite{sauls94,heffner96} for reviews),
 or they may belong to different representations of the crystal 
 point group that are either accidentally degenerate,\cite{chen93} 
 reflect a `higher  symmetry' of the crystal\cite{zhitomirsky96} or 
 of spin space\cite{machida89}.

 As briefly summarized below, experiments suggest an intriguing interplay 
 between superconductivity and antiferromagnetism; however, an interpretation 
 without significant ambiguity has not yet emerged.  Here, we focus on one
 theoretical proposal: an odd-parity superconducting state with a 
 two-dimensional order parameter that transforms like a single representation,
 the E$_2$ representation of the hexagonal symmetry group. This order
 parameter may also couple to the AFM order parameter.
 While the spatially homogeneous superconducting states of this model and 
 their response to an applied magnetic field have been studied, there is 
 comparatively little work that explores the effect of spatially inhomogeneous 
 antiferromagnetism (SIAFM) on superconductivity. Motivated by recent work of 
 Garg\cite{garg98}, which concludes that spatially varying AFM would rule out 
 the two dimensional representation models for UPt$_3$,  we examine the 
 sensitivity of an E$_{2u}$ superconductor to SIAFM through numerical 
 calculations of its response to two qualitatively different kinds of spatial
 configurations of the SIAFM.  Since the free energy functionals for an 
 E$_{2u}$ and an E$_{1g}$ superconductor are formally identical, our results 
 are also relevant to E$_{1g}$ superconductors.

 Antiferromagnetism is the prime suspect for inducing the zero-field 
 double phase transition observed in specific heat experiments on high-quality 
 crystals. A coupling between AFM and superconducting order parameters
 is suggested by a downward kink in the magnitude of the modulus of the AFM 
 order parameter at the superconducting transition.\cite{isaacs95}
 In a comparison of specific heat measurements with neutron scattering 
 experiments under pressure, Hayden {\em et al}.\cite{hayden} observed that 
 the disappearance of the double transition is correlated with the 
 disappearance of signatures of antiferromagnetism.  Recent work of 
 Keizer {\it et al.}\cite{keizer99} further supports this correlation.
 Upon substituting Pd for Pt on a small number of sites, they find that the 
 magnetic moment and the splitting of the double transition increase 
 simultaneously with increasing Pd doping.

 The nature of the antiferromagnetism is itself unusual.  A signature of an 
 AFM phase transition in thermodynamic, NMR (nuclear magnetic resonance), 
 and zero-field $\mu$SR (muon spin relaxation) experiments has so 
 far not been observed.\cite{nosignatures}  This has been taken as evidence for 
 the absence of long-range order and the existence of magnetic fluctuations on a 
 characteristic scale smaller than neutron scattering frequencies, but greater 
 than those of NMR. This temporal fluctuation has largely been ignored and the 
 antiferromagnetism has been taken to be static when considering the interaction 
 of AFM with superconductivity.  While conventional thermodynamic signatures of 
 the N\'eel temperature, $T_{N}$, have not been observed, more recent transverse 
 high-field $\mu$SR experiments have detected anomalies at $T_{N}$ as  
 identified by neutron scattering.\cite{muons2} 

\begin{figure}[b]
\noindent
\begin{minipage}{\hsize}
\epsfysize=59mm
\centerline{\rotate[r]{\epsfbox{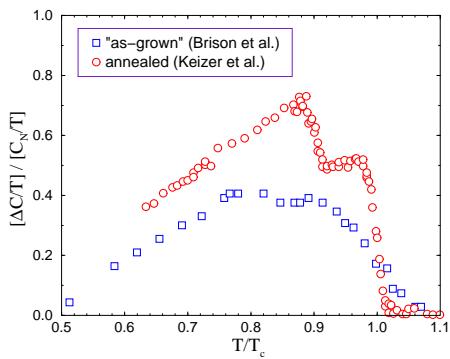}} }
\caption[]{Specific heat data of a low quality sample
showing a single broad peak\cite{spec_heat} (squares) 
compared to that for a high-quality sample.\cite{keizer99} The
latter reveals signatures of two closely spaced phase transitions.
In the normal state $C_N/T \approx 430 \,{\rm mJ/(K^2 \, mol)}$.
}
\label{shdata}
\end{minipage}
\end{figure}

 The appearance of the double transition was unexpected.  As shown 
 in Fig.~\ref{shdata}, specific heat experiments\cite{earlysh} prior to 
 1989 typically  showed a single anomalously broad peak at the
 transition to superconductivity.  An obvious explanation is that 
 AFM domains increase in size during the annealing process
 sharpening the distribution around two intrinsic superconducting 
 transitions, but X-ray\cite{isaacs95} and neutron scattering\cite{keizer99s} 
 experiments fail to show any obvious correlation between domain size and annealing.  

 A commonly held physical picture is based on an interpretation of the neutron 
 scattering data.  The AFM order with orthorhombic symmetry appears with 
 $\sim 0.02 \mu_B$ ordered moments constrained to lie in the basal plane.  
 AFM order occurs in domains of uniform 
 size $\sim 30\,$ nm, that are randomly distributed over three ${\bf q}$ vectors that 
 are oriented at $0^{\circ}$, $120^{\circ}$, and $240^\circ$ with respect to the
 ${\bf a}^*$ axis.
 The moments are essentially rigidly locked to the lattice for fields in excess of the 
 zero-temperature upper critical field of the superconducting state.\cite{lussier96,vandijk98}

 This picture is not firmly established.  Existing neutron scattering data
 is unable to rule out the possibility that instead of domains of a 
 single-${\bf q}$ structure, AFM order appears in a triple-${\bf q}$ 
 structure that preserves the symmetry of the crystal 
 lattice.\cite{vandijk98} Moreover, a recent careful analysis of the 
 neutron scattering data\cite{moreno} finds that no conclusion can be drawn 
 from existing data on whether 
 the staggered moment remains fixed to the lattice or whether it rotates 
 with an applied magnetic field.  Existing data also cannot distinguish 
 randomly oriented abutting domains with small magnetic moments from small 
 domains with magnetic moments of $\sim 1 \mu_B$ interspersed in an 
 otherwise nonmagnetic system.  

 Whether AFM occurs in a few large domains with staggered moments that are free to rotate 
 in an applied field, or whether it is strongly spatially varying with staggered moments that 
 are rigidly fixed to the lattice, is important in determining the symmetry of 
 the superconducting states, particularly for states near the upper transition temperature.  
 Several authors take the view that this quasistatic AFM acts as a symmetry breaking 
 field (SBF) and lifts the degeneracy among components of a multicomponent order parameter 
 resulting in two superconducting phase transitions separated by $\sim 50\,{\rm mK}$ 
 that are observed in zero field.\cite{heffner96,volovik88,hess89,tokuyasu90,machida89,joynt90,blount90,norman92,lukyanchuk93,park96,multi_machida} 
 A two-component odd-parity order parameter 
 that transforms like the two-dimensional $E_{2u}$ representation of the hexagonal 
 symmetry group D$_{6h}$ is one of the more promising proposals.\cite{sauls94,snote}  
 At low temperature or in the absence of the SBF, weak coupling BCS theory shows that 
 the homogeneous equilibrium state breaks time-reversal symmetry. 
 Further calculations using weak-coupling BCS theory show that thermal 
 conductivity,\cite{graf96} transverse sound attenuation,\cite{graf99} 
and upper critical
 fields\cite{choi91} of this state are in good agreement with experiments 
for temperatures in the low temperature phase.\cite{graf00} Coupling to an 
SBF has been included within a Ginzburg-Landau (GL) theory developed for a 
single-domain superconducting state.  
 Signatures of a double phase transition are apparent in the specific heat and 
lower critical field,\cite{hess89,sauls94} and in the cores of vortices.\cite{hess94} 
 In contrast to two-dimensional even-parity E$_{1g}$ and E$_{2g}$, 
and to odd-parity E$_{1u}$ order parameters, the E$_{2u}$ model can allow for 
a tetracritical point for arbitrary field 
 orientations in the H-T phase diagram.  An enhancement of this model 
 includes the competition between magnetic anisotropy and Zeeman energies 
 of the magnetic order parameter,\cite{sauls96} and reproduces the 
 angular dependence of the upper critical field observed in 
 experiment.\cite{keller94} 

 Within the E$_{2u}$ model, comparatively little has been done to explore
 the effect of coupling superconductivity and spatially inhomogeneous 
 AFM.  Motivated by the quenched domain interpretation of the 
 neutron scattering data, early work on E-representation superconductivity
 by Joynt {\it et al.}\cite{joynt90} and by Mineev\cite{mineev91}
 focused on abutting AFM domains that are uniform in size with dimensions 
 of a superconducting coherence length and considers the possibility of a 
 superconducting glass phase.  Based on a variational calculation for domains
 of uniform size and on calculations for a one-dimensional `toy model', 
 Garg\cite{garg98} argued that the pure E-representation models are
 `incompatible' with UPt$_3$ for the small domains suggested by the neutron 
 scattering experiments. 

 Taking the AFM order to be static, we explore the sensitivity
 of E$_{2u}$ superconductivity to spatially varying AFM order in 
 two models for the (disordered) domain structure of the AFM state:
 (I) abutting AFM domains of uniform size with orientations that are 
 equally distributed over the allowed ${\bf q}$-vectors, 
 and (II) small domains interspersed in a homogeneous AFM background.  
 We consider these models to represent limiting cases for the 
 configuration of the SIAFM.  
 The first model corresponds to the 
 standard interpretation of the neutron scattering data. The second model
 begins with uniform AFM order and adds `nanodomains'  
 with random position and orientation of the staggered magnetization; this 
 provides the broadening mechanism for the linewidth of neutron scattering
 data.
 We present numerical solutions of the 
 GL equations in two spatial dimensions and  
 focus on contributions to the free energy, the specific heat, and the nature 
 of the superconducting state.  Our results for Model I agree 
 with those for Garg's simple periodic model.  While neither model 
 provides an adequate account 
 of experiment, our results for Models I and II differ, and taken together
 they {\em do not} lead to Garg's conclusion that the E$_{2u}$ model is incompatible
 with UPt$_3$. Rather, they suggest that a reasonably accurate model of the spatial
 dependence of the AFM is required to make meaningful comparisons with experiment. 
 In a forthcoming work, we discuss other domain configurations along with the effects 
 of dimensionality and the possible role of superconducting glass 
 phases.\cite{grafnext}
 Central results of this paper are contained in the comparison of specific 
 heat calculations with experiment for both models.  

 In Model I, the signatures of the superconducting transitions 
 rapidly smear and broaden with decreasing domain size.  The SBF introduces 
 a convenient length scale $\xi_{\varepsilon}$ (defined below) which, for UPt$_3$,
 is some three times larger than the zero-temperature superconducting coherence 
 length.  For domain sizes of $10\xi_\varepsilon -20 \xi_\varepsilon$, 
 the calculated specific heat compares well with data from experiment.\cite{keizer99}  
 This homogeneous domain size model does not agree with neutron scattering experiments
 which, when viewed through the lens of Model I, would suggest a much smaller 
 domain size, $\sim 1 \xi_\varepsilon -3 \xi_\varepsilon$.  
 As the domain size is decreased below $\sim 2 \,\xi_\varepsilon$ 
 only a single superconducting transition appears in the specific heat.  
 While the suppression of the double transition with decreasing domain size suggests
 an obvious explanation for the appearance of the double transition upon 
 annealing of as-grown samples (see Fig.~\ref{shdata}), this explanation appears
 to be inconsistent with magnetic x-ray and neutron scattering data which are
 interpreted as showing no change in domain size with annealing.

 In contrast, Model II is not as sensitive to the density of 
 nanodomains. Specific heat signatures remain sharp even for a high density 
 of `nanodomains' and resemble those for high-quality crystals. 
 Although these signatures remain sharp, for coverages larger than 
 $\sim 75$\% only a single superconducting transition occurs.

%
%% fig 1
%

 While neither model provides an adequate account of the experimental data,
 the calculations presented here demonstrate the sensitivity
 of E$_{2u}$ superconductivity to SIAFM and suggest that a reasonably 
 accurate description of the spatial variation of the underlying AFM is crucial 
 in drawing conclusions about the symmetry of the superconducting states.

\section{Coupled Superconductivity and Antiferromagnetism}

\subsection{The E-representation model}

We begin by reviewing the Ginzburg-Landau free energy for an E$_{2u}$ 
superconductor coupled to an SBF.  We take the E$_{2u}$ gap matrix
$\Delta_{\alpha \beta}({\bf k},{\bf r})$
to be the superconducting order parameter; it has the form
\begin{equation} 
\Delta_{\alpha \beta}({\bf p}_f,{\bf r}) = 
{\bf e}({\bf p}_f) \cdot \veta({\bf r}) 
(\hat{c} \cdot \,i \vec{\sigma} \sigma_y)_{\alpha \beta} 
\ ,
\end{equation}
where ${\bf e}({\bf p}_f) = \big( e_1({\bf p}_f), e_2({\bf p}_f) \big)$
are basis functions on the Fermi surface that transform 
among each other under the operations of the D$_{6h}$ symmetry group, 
$\veta ({\bf r}) = \big( \eta_1 ({\bf r}), \eta_2 ({\bf r}) \big)$ 
are Cooper pair amplitudes that are functions of the Cooper 
pair center of mass ${\bf r}$, $\hat{c}$ is a unit vector along the 
c-axis of the crystal, and $\sigma_i$ are Pauli matrices in spin
space. Explicit expressions for E$_{2u}$ and other basis functions 
may be found in, {\em e.g.}, Refs.~\onlinecite{sauls94,heffner96} 
and~\onlinecite{graf96}.
The GL free energy density is  
\begin{equation} \label{f_e2u}
f({\bf r}) = f_{\rm bulk}({\bf r}) + f_{\rm grad}({\bf r}) 
 + f_{\rm field}({\bf r}). 
\end{equation}
Symmetry constrains the form of 
the bulk, gradient, and field contributions 
for an $E_{2u}$ superconductor to be\cite{sauls94}
\begin{eqnarray}
& f_{\rm bulk}  =  \alpha(T) |\veta|^2 + \beta_1 |\veta|^4 + 
 \beta_2  |\veta \cdot \veta|^2 \ , &\\
& f_{\rm grad}  =  \kappa_1 (D_i \eta_j)(D_i \eta_j)^* +
 \kappa_2 (D_i \eta_i)(D_j \eta_j)^* &\nonumber\\
& \quad + 
 \kappa_3 (D_i \eta_j)(D_j \eta_i)^* +
 \kappa_4 (D_z \eta_j)(D_z \eta_j)^* \ ,&\\
& f_{\rm field}  =  \frac{1}{8\pi} |{\bf b}|^2 \ , &
\end{eqnarray}
where $\alpha(T) = \alpha_0 \, (T-T_0)$, 
$\alpha_0$ is a constant,
$T_0$ is the transition temperature,
${\bf b} = \mbox{\boldmath$\partial$} \times {\bf A}$ is the magnetic field,
and 
${D}_j = \partial_j - i (2 e/\hbar c){A}_j$ 
is the gauge-invariant gradient. 
In the calculations that follow for a spatially 
varying SBF, we restrict ourselves to two dimensions and
$\kappa_4$ plays no role.  In weak-coupling BCS theory
for an E$_{2u}$ order parameter, the parameters 
$\kappa_2$ and $\kappa_3$ are small for Fermi surfaces with axial 
symmetry;\cite{sauls94} we take $\kappa_2 = \kappa_3 = 0$ and 
write $\kappa \equiv \kappa_1$ for notational convenience.  Weak-coupling 
BCS theory also predicts that $\beta_2 = \beta_1/2$ independent of the shape 
of the Fermi surface and that the homogeneous equilibrium order parameter 
that minimizes the free energy is doubly degenerate, breaks time-reversal 
symmetry, and is of the form $\veta \sim (1, \pm i)$. 

The orientation of the AFM order parameter ${\bf N}({\bf r})$ may fluctuate 
dynamically.  An estimate of the magnetic fluctuation time $\tau_{\rm mag}$, from 
the energy-resolution limited magnetic Bragg peaks, obtained from 
elastic neutron scattering experiments, 
gives $\tau_{\rm mag} \sim 500 \, {\rm ps}$.\cite{aeppli94}
Because this fluctuation time is slow compared to the characteristic time scale of 
the superconducting state $\tau_{sc} \sim h/\Delta_0 \sim 50\, {\rm ps}$, 
we take the SBF to arise from static AFM order\cite{joynt90} and calculate 
equilibrium solutions of the GL functional in the presence of 
spatially varying AFM order.

The transition temperature below which AFM order occurs is an order of 
magnitude larger than the superconducting transition.  In a mean-field description
of the AFM order, the order parameter is well developed at the transition to 
superconductivity. Below the superconducting 
transition, the neutron scattering data suggests
that the modulus of the staggered magnetization decreases. Over the temperature 
range for which a GL theory of superconductivity is valid, the 
modulus of the staggered magnetization changes by less than $5$\%. 
Since this change is small, we take the staggered magnetization to be a fixed
external field and neglect the effect of superconductivity on AFM
order.\cite{footnote_AFM} 
The leading contribution to the free energy from the coupling 
of AFM and superconductivity is second order in $\veta$ and of the form
\begin{eqnarray}\label{f_sbf}
\displaystyle
f_{\rm sbf} &=& 2 \varepsilon \, \alpha_0 \,
\displaystyle
\left( | {\bf N} \cdot \veta |^2 - {1\over 2}|\veta|^2 \right) 
\nonumber\\ &=& 
 \varepsilon \, \alpha_0 \, \veta^\dagger
 \left(
 \begin{array}{cc}
 -\cos\,2\theta & \sin\,2\theta \\
 \sin\,2\theta  & \cos\,2\theta 
 \end{array}
 \right)  \veta
\ , 
\end{eqnarray}
where ${\bf N} = \big( \sin\,\theta({\bf r}), \cos\,\theta({\bf r}) \big)$ is a
normalized direction vector of the staggered magnetization, 
and $\varepsilon$ is a coupling constant
proportional to the square of the staggered magnetization.
Symmetry also allows a coupling of the SBF to the gradient in the E$_{2u}$ 
model,\cite{sauls94,sauls94a} so that $f_{\rm grad}$ is given by
\begin{equation}
f_{\rm grad} = \kappa_1^+ |D_i \eta_1|^2 + \kappa_1^- |D_i \eta_2|^2 
\end{equation}
with 
\begin{equation}
 \kappa_1^\pm = \kappa_1 ( 1 \pm \varepsilon_{\perp} {\bf N}^2).
\end{equation}
This term, together with the $\hat{c}$ axis gradient terms, determines the
kink and possible tetracritical point on the upper critical field curve. Because
the magnitude of the gradient coupling to the SBF is small, being of the order
$\varepsilon_{\perp} \sim \varepsilon \sim \Delta T_c / T_c$, it will not 
significantly affect our results, and so we neglect the (direct) coupling of ${\bf N}$ 
to ${\veta}$ through the gradient terms for the calculation of thermodynamic properties 
in zero magnetic field. 

The symmetry breaking term in Eq. (\ref{f_sbf}) is combined with Eq. (\ref{f_e2u})
to give the free-energy density 
\begin{eqnarray}\label{gl_eq}
f &=& \alpha_{-}(T) |\eta_1|^2 + \alpha_{+}(T) |\eta_2|^2
+ \varepsilon\alpha_0\sin\,2\theta \,(\eta_1 \eta_2^* + c.c.) 
\nonumber\\&&
+ \beta_1 ( |\eta_1|^2 + |\eta_2|^2 )^2
+ \beta_2 | \eta_1^2 + \eta_2^2 |^2
\nonumber\\&&
+  \kappa (D_i \eta_j)(D_i \eta_j)^*
+ \frac{1}{8\pi} {\bf b}^2 \; ,
\end{eqnarray}
with
$\alpha_{\pm}(T) = \alpha(T) \pm \varepsilon\alpha_0\cos\,2\theta$.
For temperatures very near the normal-superconducting transition, 
the second order terms that include the coupling to the SBF dominate 
and `real' phases of the form 
${\veta} \sim (\cos \theta \, , \sin \theta) e^{i\varphi}$, 
minimize the free energy.

\subsection{Single AFM Domain and Estimation of GL Parameters}

Without loss of generality, the salient features of coupling 
to uniform AFM (single infinite domain) can be seen by taking
${\bf N}({\bf r}) = (0, 1)$ and the coupling to 
the SBF to be positive, $\varepsilon > 0$, so that $f_{\rm sbf}$ favors 
${\veta} \perp {\bf N}$.  At the temperature 
$T_{c+}^{\rm hom} = T_0 + \varepsilon$, a phase transition
occurs from the normal state to a spatially homogeneous 
superconducting phase ${\veta} \propto (1,0)$. At a 
lower temperature $T_{c-}^{\rm hom} = T_0 - \varepsilon/\beta$,
there is a second phase transition to a time-reversal symmetry 
breaking phase $\veta \propto (1, \pm i r(T))$, where 
$\beta = \beta_2/\beta_1$ and $r(T)$ is a function that
grows rapidly and smoothly from $0$ to $1$ as $T$ is lowered.\cite{hess89}

The specific heat jumps (per volume)
at the two phase transitions $T_{c+}^{\rm hom}$
and $T_{c-}^{\rm hom}$, 
measured relative to the normal state, 
are calculated from a derivative of 
the free energy and are given by\cite{hess89}
\begin{eqnarray}
\triangle C_{+}^{\rm hom} & = & 
{ T_{c+}^{\rm hom} \, \alpha_0^2}/{ 2\beta_{12} } \ , 
\\
\triangle C_{-}^{\rm hom} & = & 
{ T_{c-}^{\rm hom} \, \alpha_0^2 }/{ 2\beta_1 } \ ,
\end{eqnarray}
with $\beta_{12} = \beta_1 + \beta_2$.
The ratio of the heat capacity jumps is
\begin{equation}
\frac{ \triangle C_{-}^{\rm hom} }{ \triangle C_{+}^{\rm hom} }=
\frac{T_{c-}^{\rm hom}}{T_{+}^{\rm hom}} ({1 + \beta}) \ .
\end{equation}
Using these equations and the specific heat data, we 
estimate the ratio $\beta \approx 0.42 - 0.65$.\cite{spec_heat,keizer99}
Noting that the weak-coupling value of $\beta$ lies comfortably 
in this range, we take $\beta=1/2$ in the calculations below. 
Further, the measured $\hat{c}$ axis $H_{c2}$ slopes\cite{adenwalla90,bruls90} of 
$\sim 6.6\,  {\rm T/K}$ imply
within a coarse-grained analysis that
$\kappa/\alpha_0 \approx 50\,  {\rm K\,nm^2}$,
i.e., a zero-temperature GL coherence length
$\xi_0 = \sqrt{ \kappa/(\alpha_0 T_{c+}^{\rm hom})} \approx 10\,{\rm nm}$.
This estimate of $\xi_0$ in the basal plane is in very good agreement
with other reports.\cite{kleiman92}

\section{Spatially Inhomogeneous Antiferromagnetism}

 We now consider the possibility that the orientation of ${\bf N}$ 
varies in the crystal lattice, ${\bf N} = {\bf N}({\bf r})$, and explicitly 
investigate two cases: 
(1) abutting antiferromagnetic domains of uniform size, $\xi_{\rm afm}$, with 
orientations distributed equally among the three possible ${\bf q}$-vectors, and 
(2) randomly dispersed `nanodomains'  with 
characteristic dimensions of order the superconducting coherence length. 
For SIAFM, an analytic solution is generally no longer 
possible and it is necessary to solve the GL equations numerically.  The resulting 
superconducting state will be complicated, because of the competition between the 
condensation energy gained by $\veta$ orienting in directions preferred by the SBF and the 
gradient energy cost to twist the orientation of the order parameter from domain 
to domain.  The response of an E$_{2u}$ superconductor to these two models for
SIAFM differs as described below.

\subsection{Preliminaries}

For numerical calculation, it is convenient to introduce 
a scaled order parameter $\tilde{\veta} = \veta / \eta_0$, 
where ${\eta_0}$ is the modulus of the real phase solution of a 
homogeneous single-domain,
$\eta_{0} = \sqrt{ \alpha_0 \varepsilon t/ 2 \beta_{12}}$, 
that appears in the presence of uniform
AFM order at a transition temperature 
$T_{c+}^{\rm hom} = T_{0} + \varepsilon$.  
All temperatures are given in terms of a reduced temperature
$t = (T_{c+}^{\rm hom} - T)/ \varepsilon$.
Scaling Eq.~(\ref{gl_eq}) to the magnitude of the 
free energy density
of the homogeneous single-domain solution in the high-temperature phase,
$|f_{0}(t)| =  (\alpha_0 \varepsilon t)^2 / (4 \beta_{12})$, 
a dimensionless Ginzburg-Landau free energy density, $\tilde{f}$, 
is obtained which is of the form 
\begin{eqnarray}
\tilde{f} & = & -2 \Big[ 
 (1 - \tau {\sin^2 \theta} ) | \tilde{\eta}_1 |^2 +
 (1 - \tau {\cos^2 \theta} ) | \tilde{\eta}_2 |^2 
  \nonumber \\ && 
  -\tau {\sin 2\theta} \,{\rm Re\,} \tilde{\eta}_1 \tilde{\eta}_2^*
  - {1\over 2} |\tilde{\veta}|^4 + {{2\beta} \over {1 + \beta}} 
  | {1\over 2} \tilde{\veta} \times \tilde{\veta}^* |^2
\Big] 
  \nonumber \\ &&
  +\tau ( \, |\tilde{D}_i \tilde{\eta}_1|^2 + 
  |\tilde{D}_i \tilde{\eta}_2|^2 \,) 
 \, , 
\end{eqnarray}
where all lengths are measured in terms of the SBF 
length $\xi_\varepsilon = \sqrt{\kappa / \alpha_0 \varepsilon}$, 
$\tilde{D}_i = \xi_\varepsilon D_i$, and $\tau = 2/t$.

For ease of calculation, we take a square computational mesh with a step 
size $\triangle x = \triangle y = 0.2\,\xi_\varepsilon$.  We tested our GL simulations 
on triangular lattices, which are more natural given the apparent hexagonal crystal
symmetry of UPt$_3$, and have found no qualitative differences for averaged quantities.  
Periodic boundary conditions were imposed.  We take the GL parameters 
to be $\beta=1/2$ and $\kappa = \alpha_0 \varepsilon \xi_\varepsilon^2 $.

As the superconducting order parameter twists to accommodate the 
spatially inhomogeneous symmetry breaking field, time-reversal 
symmetry breaking phases may appear in localized regions even for 
temperatures near the normal-superconducting phase boundary where 
`real' phases are expected to dominate.  To detect the appearance
of these phases, we calculate 
\begin{equation}
{\bf M}_{\rm orb}({\bf r})  =  \, {{1} \over {2 i}} \;
 \tilde{\veta}({\bf r}) \times \tilde{\veta}({\bf r})^* \ ,
\end{equation}
which is (apart from a factor dependent on the gradient terms\cite{sauls94})
the spontaneous magnetization that arises from the
internal orbital motion of a Cooper pair. Note that
${\bf M}_{\rm orb}$ is a real vector constrained by symmetry to point along 
$\pm \hat{c}$.
For a homogeneous single domain,  ${\bf M}_{\rm orb}$ vanishes between 
the two transitions, as it must for a `real' phase.  At the lower 
transition, the second component of the order parameter begins to
grow with a phase relative to the first and ${\bf M}_{\rm orb}$ 
increases rapidly with decreasing temperature signaling 
broken time-reversal symmetry.  As expected, the temperature dependence 
of ${\bf M}_{\rm orb}$ is consistent with a second order phase transition
in mean field theory, 
$|{\bf M}_{\rm orb}(T)| \sim \sqrt{T_{c-}^{\rm hom} - T}$,
for $T \leq T_{c-}^{\rm hom}$.

The calculation of free energies and spatially averaged quantities
involves a summation over the lattice.  We adopt the notation
\begin{equation}
\langle A ({\bf r}) \rangle_{\bf r} = N^{-2} \; \sum_{i=1}^N
\sum_{j=1}^N \; A(x_i, y_j) \, ,
\end{equation}
where $N^2$ is the number of computational mesh points
and the position vector ${\bf r} = (x,y)$. 
In this notation, the spatially averaged orbital moment is 
$\langle \bf{M}_{\rm{orb}}({\bf r}) \rangle_{\bf r}$.
In the case of Model I, we solved for several antiferromagnetically ordered
domain sizes on a square lattice.  For Model II we have found solutions of the
GL equations for different sets of random configurations of the SBF and have 
not found any discernible differences for spatially averaged quantities. 
We attribute this
`self-averaging' of the quenched disorder of the randomized SBF to our 
relatively large system size (this would be exact for an infinite system) 
and to the large concentration of nanoscale defects ($9\% - 79\%$).

%
%% fig 2
%

\begin{figure}
\noindent
\begin{minipage}{\hsize}
\epsfxsize=45.0mm
\centerline{{\epsfbox{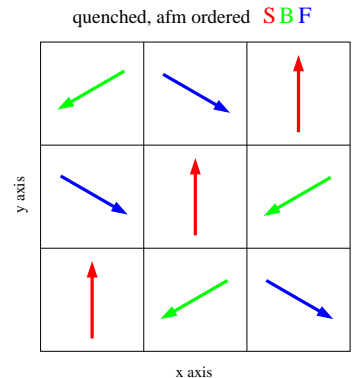}} }
\vspace*{0.1in}
\caption[]{Structure of AFM domains assumed in Model I.
For convenience a computational mesh with a square geometry 
was chosen; explicit comparison with calculations on triangular
meshes do not change our central conclusions.
}
\label{m1-domains}
\end{minipage}
\end{figure}

\subsection{Model I - Checkerboard}

The spatially varying staggered moment configuration consists of abutting 
domains as shown in Fig.~\ref{m1-domains}.  The orientation pattern of 
${\bf N}({\bf r})$ was generated so that nearest-neighbor domains do 
not have an SBF of the same orientation and so that there is no net staggered 
magnetization.  The influence of the SBF is greatest at high temperature and 
the nature of the superconducting phase depends on the size of the AFM domains 
relative to the superconducting coherence length.
At sufficiently low temperature, the order parameter is 
${\veta}({\bf r}) \propto (1,i)$ 
to an excellent approximation; for most purposes the coupling to the SBF leads 
to a negligible correction to the order parameter.  

\begin{figure}
\noindent
\begin{minipage}{\hsize}
\epsfysize=70mm
\centerline{\rotate[r]{\epsfbox{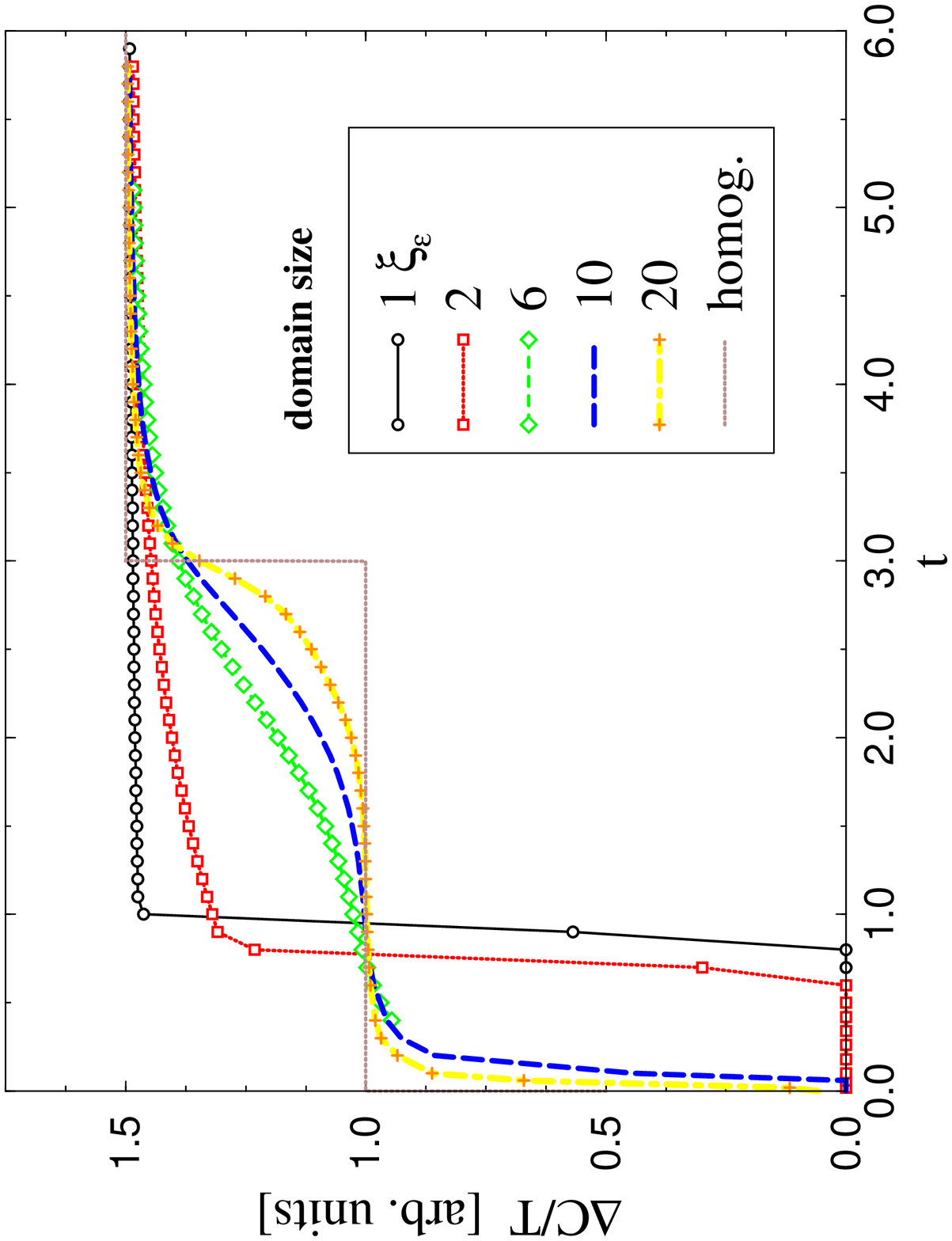}} }
\vspace*{0.1in}
\epsfysize=70mm
\centerline{\rotate[r]{\epsfbox{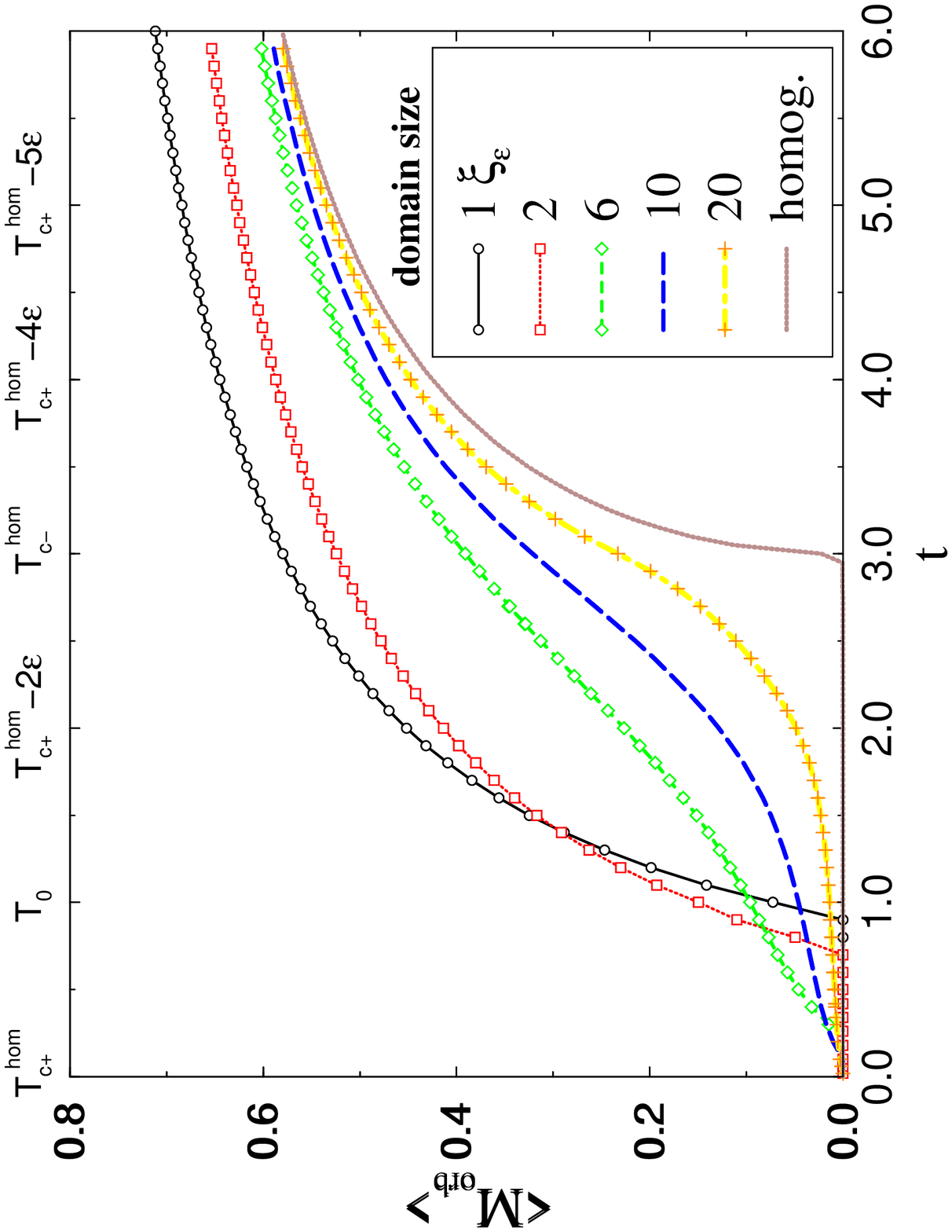}} }
\nopagebreak
\caption[]{
Top: The specific heat for Model I
in units of the upper specific heat jump
of the spatially homogeneous system, 
$\triangle C_{+}^{\rm hom}/T_{c+}^{\rm hom}$,
for various AFM domain sizes measured in 
units of the SBF length $\xi_\varepsilon$.
The temperature $t$ is measured relative to the upper superconducting 
transition and {\em decreases} in the positive x-direction,
$t = (T_{c+}^{\rm hom} - T)/\varepsilon$.
Bottom: The corresponding spatially averaged 
spontaneous magnetization 
$\langle {\bf M}_{orb}({\bf r}) \rangle_{\bf r}$
as a function of $t$ and for the same domain sizes. 
}
\label{m1-results}
\end{minipage}
\end{figure}

The effect of domain size on the superconducting transitions is apparent in 
Fig.~\ref{m1-results}.  For large domains two reasonably sharp transitions
are apparent.  They smear rapidly as the domain size is decreased until
for $\xi_{\rm afm} \sim 2 \,\xi_\varepsilon$ only one transition appears for the size
of our computational mesh together with imposed periodic boundary conditions.
A comparison of the numerical calculations of Fig.~\ref{m1-results} for our
model in two dimensions with the specific heat calculations for Garg's 
one-dimensional model presented in Fig.~2 of Ref.~\onlinecite{garg98} 
shows that these are in good agreement given the simplicity of the `toy' 
model.\cite{gcomp}

%
%% fig 3 (a-b)
%

The lower panel of Fig.~\ref{m1-results} shows the temperature dependence of the
spatially averaged spontaneous magnetization which reflects the  nature of the 
two phase transitions.  
For an infinitely large single domain, $\langle {\bf M}_{\rm orb}({\bf r}) \rangle_{\bf r}$ 
vanishes in the high temperature phase where the order parameter is 
`real,' and rapidly increases below the lower phase  transition, 
$T^{\rm hom}_{c-}$, as described above.
As the domain size decreases,
$\langle {\bf M}_{\rm orb}({\bf r}) \rangle_{\bf r}$ becomes finite but remains small for 
temperatures below $T_{c+}$ and above $T^{\rm hom}_{c-}$.  The time-reversal symmetry 
breaking phase appears with increasing strength as the order parameter tries 
to twist from domain to domain. 
For the largest domains, the 
specific heat shows two phase transitions (albeit rounded by spatial fluctuations), 
even though $\langle {\bf M}_{\rm orb}({\bf r}) \rangle_{\bf r}$ is finite between 
the two transitions.  For the smallest domain sizes only one transition is 
apparent\cite{smallcomp}
with an onset approximately that of
the transition temperature in the absence of coupling to the SBF, $T_0$.

%
%% fig 4
%

\begin{figure}
\noindent
\begin{minipage}{\hsize}
\epsfysize=80mm
\centerline{{\epsfbox{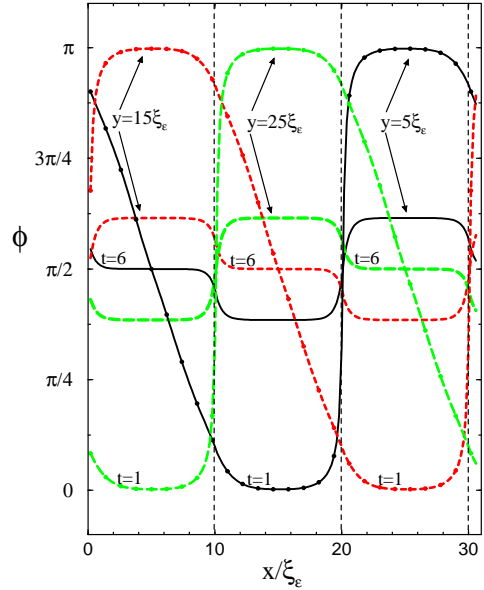}} }
\caption[]{
Spatial cuts of the relative angle $\phi$ 
between the order parameters $\eta_1$ and $\eta_2$ 
along the $x$ axis for a fixed $y$-coordinate
at temperatures $t=1$ and $t=6$.
The AFM domain size is $10\, \xi_\varepsilon$ and
$\phi = \angle(\eta_1, \eta_2)$.
The same parameters are used in Fig.~\ref{m1-scds}.
A relative angle of $\phi = \pi/2$ signals a superconducting
phase that breaks time-reversal symmetry.
The low-temperature phase ($t=6$) breaks time-reversal
symmetry on average, while the high-temperature phase
($t=1$) breaks time-reversal symmetry predominantly in 
the domain walls (indicated as vertical dashed lines).
}
\label{cuts}
\end{minipage}
\end{figure}

\begin{figure}
\begin{center}
\noindent
\begin{minipage}{0.99\textwidth}
\epsfysize=75mm
\centerline{{\epsfbox{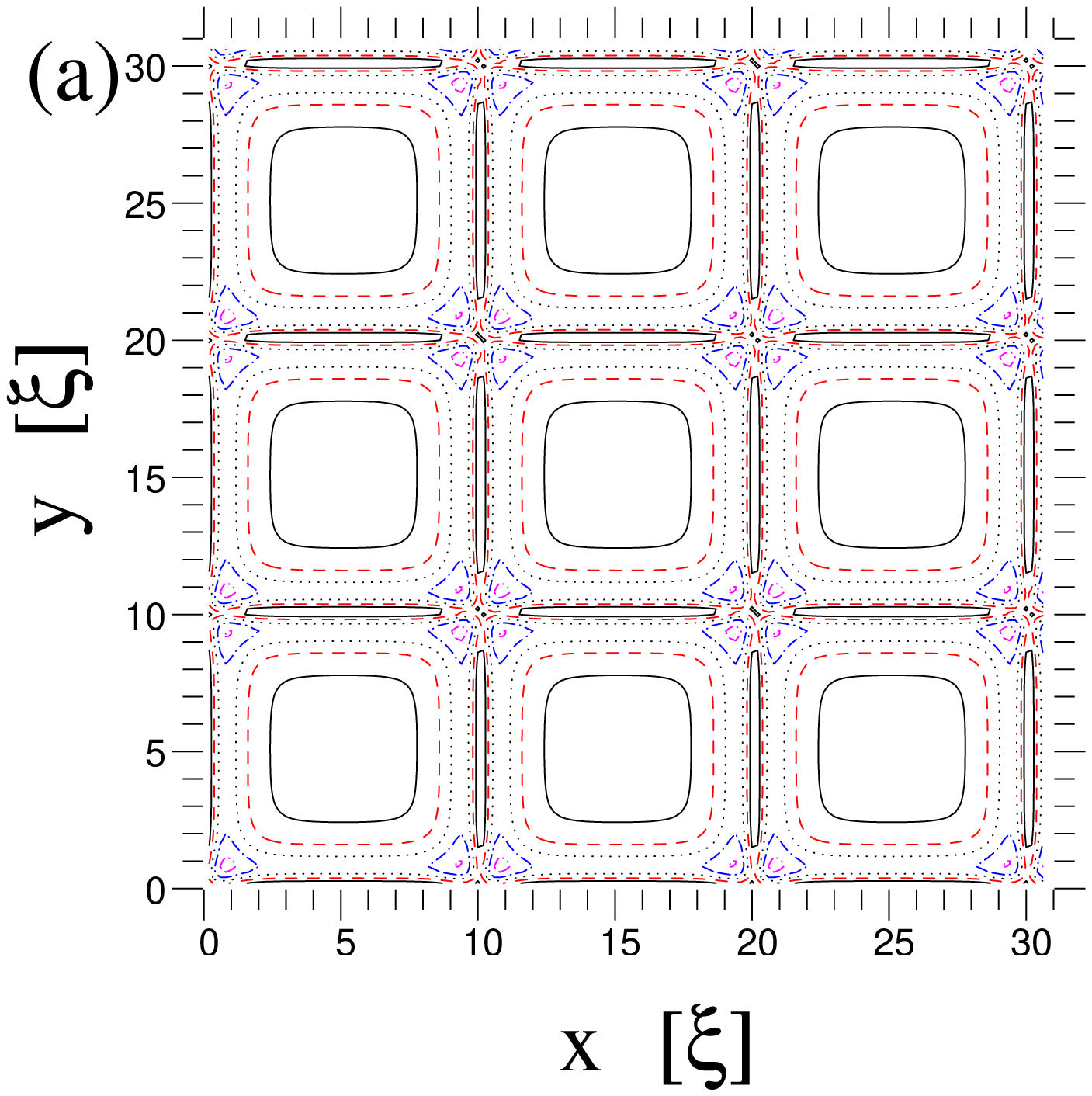}} 
\epsfysize=75mm
\hfill {\epsfbox{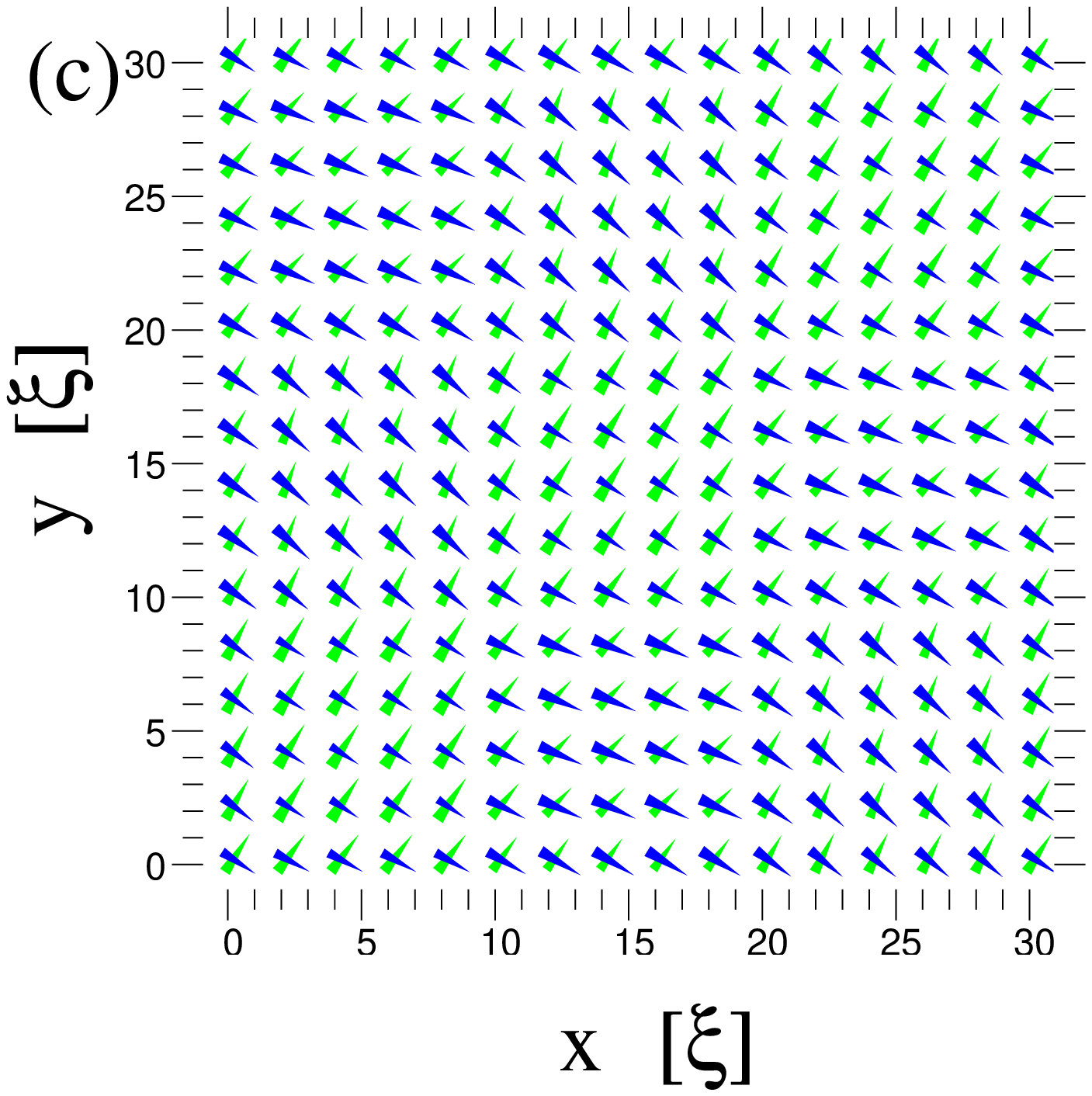}} }
\end{minipage}
\noindent
\begin{minipage}{0.99\textwidth}
\epsfysize=75mm
\centerline{{\epsfbox{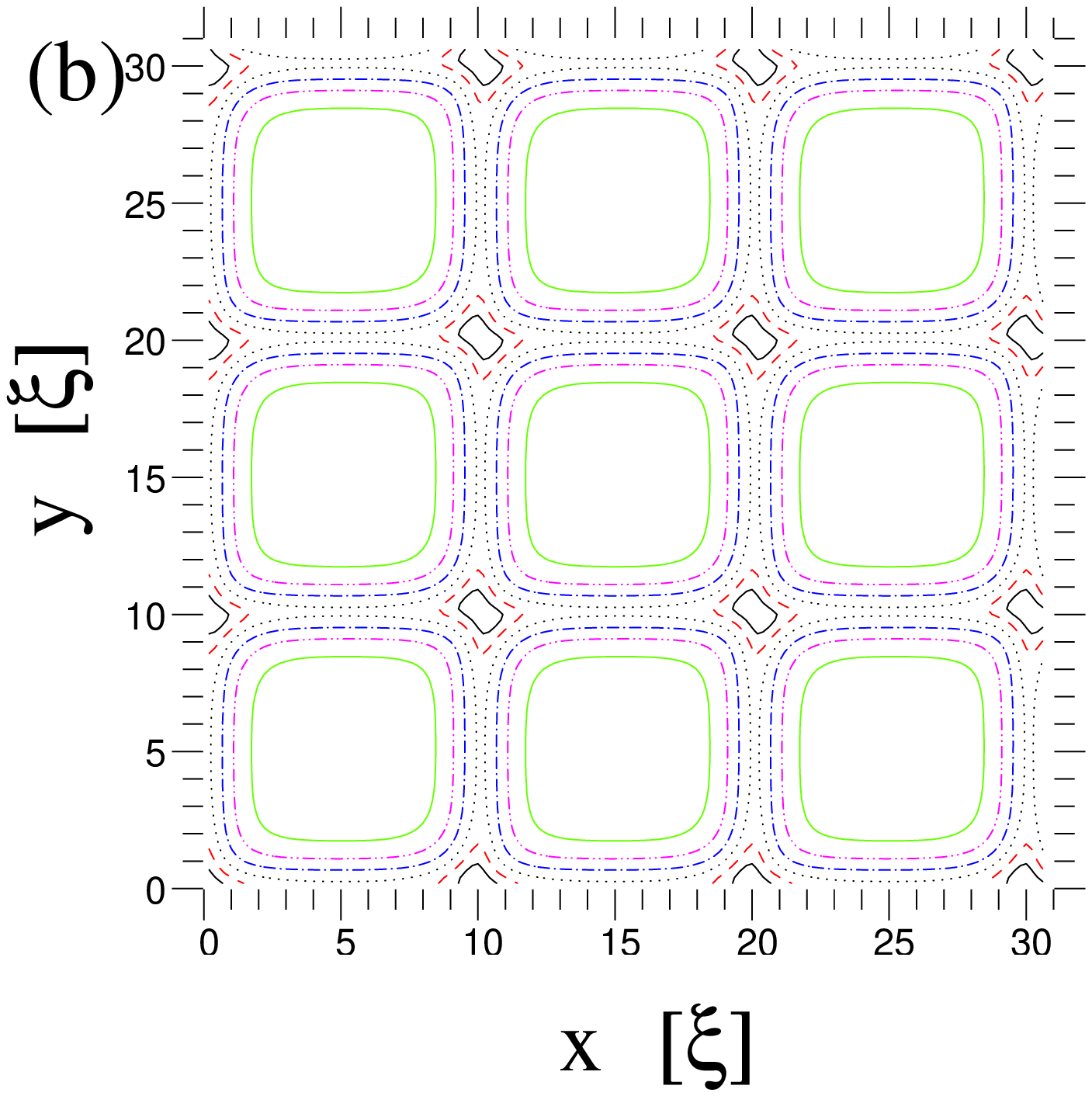}} 
\epsfysize=75mm
\hfill {\epsfbox{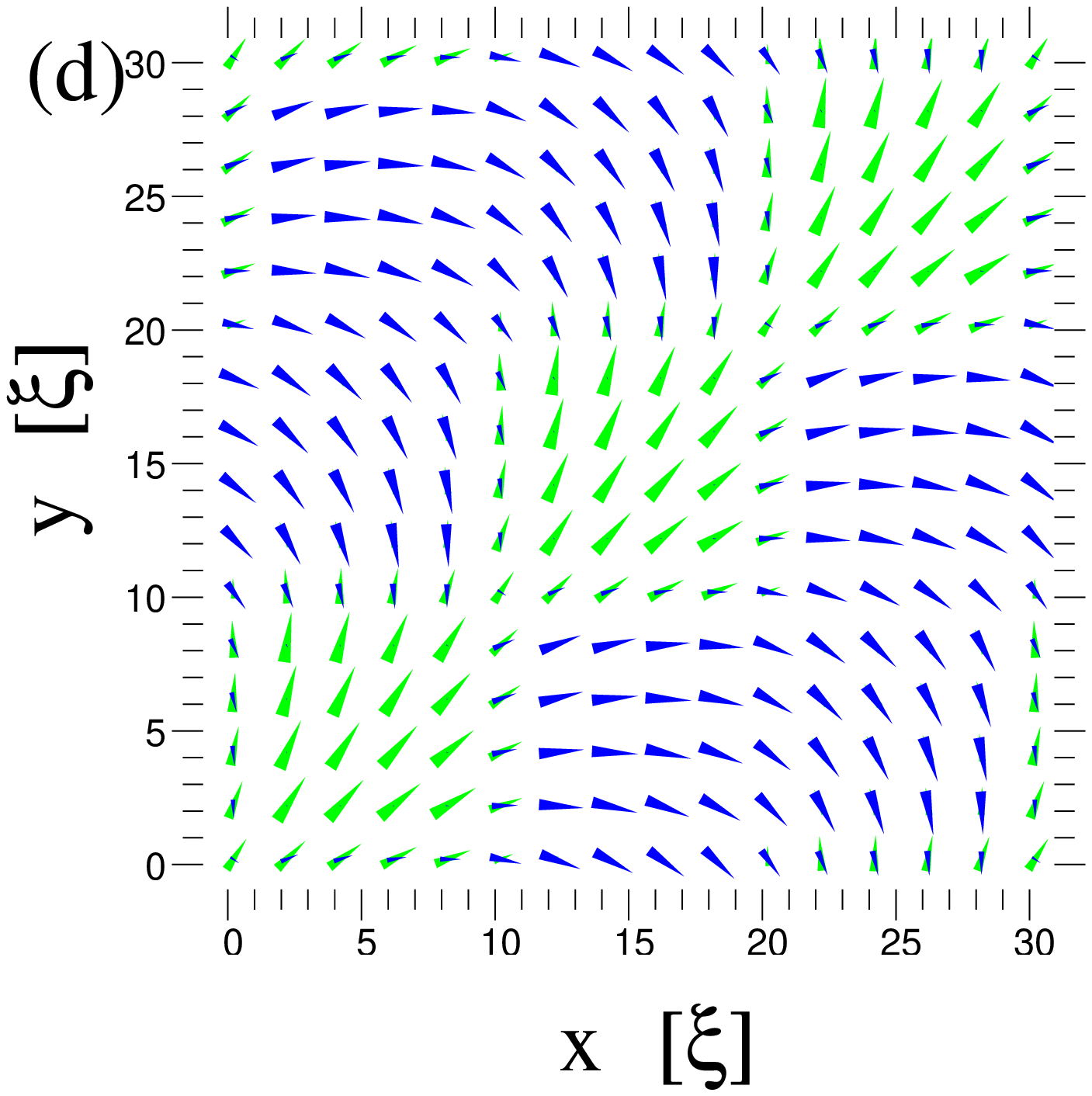}} }
\end{minipage}
\end{center}
\vspace*{0.1in}
\caption[]{Domain structure of the Model I superconducting state 
induced by the coupling to homogeneous AFM domains of size 
$10 \, \xi_\varepsilon \times 10 \,\xi_\varepsilon$ with three equivalent orientations,
$0^\circ, \pm 120^\circ$, of the SBF.  
Distances are in units of $\xi_\varepsilon$.
Panel (a): Contour plot of 
$|\veta({\bf r})|/\langle |\veta({\bf r})| \rangle_{\bf r}$ at steps 
0.999 (solid), 1.000 (dash), 1.001 (dot), 1.002 (dash-dot), and
1.003 (dash-dot-dot) at temperature $t = 6$. $|\veta|$ is minimum 
in the domain center.
Panel (b): Contour plot of 
$|\veta({\bf r})|/\langle |\veta({\bf r})| \rangle_{\bf r}$ at steps 
0.80 (solid), 0.85 (dash), 0.90 (dot), 0.95 (dash-dot),
1.00 (dash-dot-dot), and 1.05 (light solid)
at temperature $t = 1$. $|\veta|$ is maximum in the domain center.
Panels (c-d): 
The complex order parameter components $\eta_1$ (\green{light})
and $\eta_2$ (\blue{dark}) are plotted as 2D vectors, where the relative 
size of the vector is proportional to its magnitude.
In the low temperature phase, $t=6$, in panel (c), 
$\eta_1 \perp \eta_2$ on average, while in the high temperature phase
in panel (d), $\eta_1 || \eta_2$ almost always.
The orientation of the SBF is the same as in Fig.~\ref{m1-domains}.
}
\label{m1-scds}
\end{figure}

The twisting and flapping of the two superconducting order parameters $\eta_1$ and
$\eta_2$ across the domain walls is apparent in the plots of the relative phase 
angle $\phi$ between $\eta_1$ and $\eta_2$ in Figs.~\ref{cuts} and \ref{m1-scds}. 
The flapping or unwinding of the relative phase in Fig.~\ref{cuts}
follows the AFM on average, producing reasonable narrow 
(order $\sim \xi_\varepsilon$) superconducting domain walls in registry
with the AFM domain walls.  Occasionally, the relative phase unwinds from 
$0$ to $\pi$ over an entire domain.
Fig.~\ref{m1-scds} shows the spatially varying superconducting order parameter
for the particular AFM
configuration shown in Fig.~\ref{m1-domains} and a domain size of $10 \xi_\varepsilon$. 
The spatial variation of the modulus of $\veta$ shown 
at low and at high temperature in Fig.~\ref{m1-scds},
panels (a) and (b), tracks the underlying AFM domain 
structure. As ex-

\end{multicols}

%
%% fig 5 (a-d)
%

\lrule
\begin{multicols}{2}

\noindent
pected, at high 
temperature, there is a suppression of superconductivity at the domain walls.
At low temperature, there is a small reduction in the component of ${\veta}$
parallel to ${\bf N}$ in the $\sim (1,i)$ phase preferred by the fourth order terms 
in the free energy and a negligibly small suppression of superconductivity 
occurs in the ${\em center}$ of the domain.
At high temperature, the components of the superconducting order parameter 
are `real' in the interior of the domains. The orientation of the order 
parameter attempts to follow the AFM order on average.
The orientation of ${\veta}$ does not follow that of ${\bf N}$ 
perfectly.
A side-by-side comparison of Fig.~\ref{m1-scds}(d) in comparison with 
Fig.~\ref{m1-domains} shows that even for these relatively large domains, 
the different orientation of the SBF in adjacent domains forces a compromise in 
the orientation of $\veta$ as it twists from
domain to domain.  For example, $\veta$ makes a $45^{\circ}$ angle with respect to
the SBF at the center of the domain $(0-10,0-10)$ in Fig.~\ref{m1-scds}(d).
For temperatures below the second phase transition, the superconducting phase is 
essentially $(1, i)$ with a small perturbation from the AFM order. 
This is evident from Figs.~\ref{cuts} and \ref{m1-scds}(c) 
which show a sizeable phase 
angle $\sim \pi/2$ between the $\eta_1$ and $\eta_2$ components. 
At high temperature, the order parameter is mildly suppressed at the 
domain walls where the relative phase angle between the $\eta_1$ and 
$\eta_2$ components is sizeable, reaching $\sim \pi/2$ at the corners, 
indicative of the appearance of time-reversal symmetry breaking phases 
in the domain walls.

%
%% fig 6
%

\begin{figure}
\noindent
\begin{minipage}{\hsize}
\epsfysize=75mm
\centerline{{\epsfbox{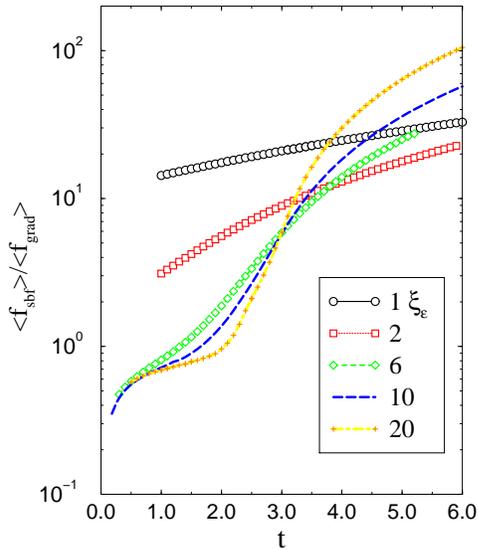}} }
\caption[]{The ratio of the contributions to the free energy from
the SBF coupling term and the gradient term,  
$\langle f_{\rm sbf}({\bf r}) \rangle_{\bf r}/\langle f_{\rm grad}({\bf r})
\rangle_{\bf r}$,
for Model I and for various domain sizes.  Note that for the smallest
domains the ratio is significantly larger at high temperatures, $t < 3$,
reflecting a new 
(strongly disordered) state in which the twisting of the order parameter
decouples from the AFM order and on average 
$\langle \veta({\bf r}) \rangle_{\bf r} \sim (1,i)$.
}
\label{m1-ratio}
\end{minipage}
\end{figure}

For various domain sizes, the nature of the spatially inhomogeneous 
superconducting state is reflected in the ratio $\langle f_{\rm sbf}({\bf r})
\rangle_{\bf r}/
 \langle f_{\rm grad}({\bf r}) \rangle_{\bf r}$, where 
$\langle f_{\rm sbf}({\bf r}) \rangle_{\bf r}$ measures, in an average way, the
extent to which the superconducting order parameter tracks the twisting of 
the SBF, and $\langle f_{\rm grad}({\bf r}) \rangle_{\bf r}$ measures the 
energy cost of twisting the order parameter.   
The ratio $\langle f_{\rm sbf}({\bf r}) \rangle_{\bf r}/ 
 \langle f_{\rm grad}({\bf r}) \rangle_{\bf r}$ plotted in Fig.~\ref{m1-ratio} 
is largely temperature independent and insensitive to domain size near 
$T_{c+}$, {\em i.e.}, $t\to 0$, for large domains. For the smallest domain size, 
the ratio is significantly larger owing mostly to a smaller 
$\langle f_{\rm grad}({\bf r}) \rangle_{\bf r}$ and the stiffness of the 
condensate. This suggests the appearance of a qualitatively different 
phase -- a strongly disordered superconducting phase.\cite{glassnote}  
As can be seen from Fig.~\ref{m1-results},
near $T_{c+}$ this phase differs from that for larger domains, because 
time-reversal symmetry is broken globally and not just in the spatially restricted 
regions between superconducting domains.

\begin{figure}
\noindent
\begin{minipage}{\hsize}
\epsfysize=75mm
\noindent
\centerline{\rotate[r]{\epsfbox{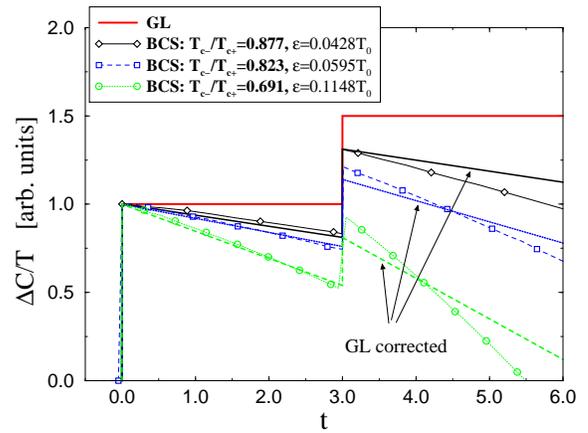}} }
\caption[]{
The specific heat for an $E_{2u}$ superconductor 
computed from weak-coupling BCS theory for various 
SBF coupling strengths $\varepsilon$ (symbols), and from GL theory
for a single AFM domain. The BCS theory temperature 
dependence can be approximately recovered from the GL theory
specific heat using a multiplicative factor 
that is linear in $t \varepsilon$.
}
\label{bcs-gl}
\end{minipage}
\end{figure}

Since GL theory does not predict the proper temperature dependence of the specific
heat, we apply a correction to recover the correct temperature dependence 
near the double transitions so that GL results may be compared
with experiment.  We compute the specific heat self-consistently
within weak-coupling BCS theory for a homogeneous single AFM domain for 
different sets of $T_c$ splittings in the clean limit and compare it with
the GL result. A reasonable approximation to the BCS theory temperature dependence 
for a small temperature range below the normal-superconducting phase transition
can be obtained by multiplying the homogeneous GL result by a
factor proportional to $\varepsilon t$, as shown in Fig.~\ref{bcs-gl}.

%
%% fig 7
%

\begin{figure}
\noindent
\begin{minipage}{\hsize}
\epsfysize=75mm
\centerline{\rotate[r]{\epsfbox{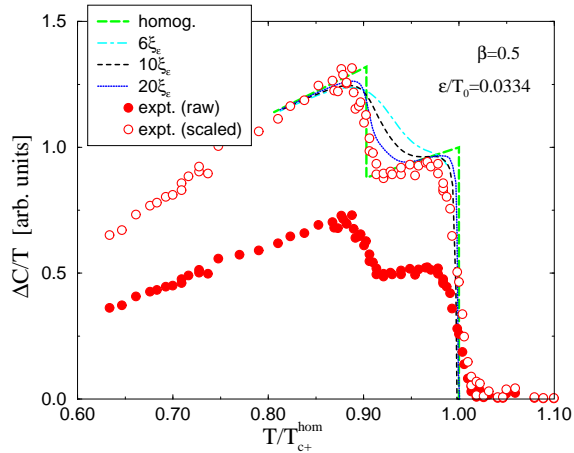}} }
\caption[]{Calculated specific heat for Model I
for a single domain and for various domain sizes
in comparison with experiment 
(filled circles) normalized by $C_N/T$.\cite{keizer99} 
For ease of comparison, the experimental data scaled by 
a numerical factor (open circles) is also presented.
The $T$-dependence of the GL results has been corrected
according to Fig.~\ref{bcs-gl}.
}
\label{fig:central}
\end{minipage}
\end{figure}

 Our results are shown together with specific heat experiments in 
Fig.~\ref{fig:central}.
For ease of comparison, the experimental results have been scaled so that the peak
in the specific heat at the lower phase transition agrees with the corresponding
feature for a homogeneous SBF.  Calculations for domain sizes 
$\sim 6 \xi_\varepsilon - 20 \xi_\varepsilon$ show two phase transitions that are 
blurred by the spatial inhomogeneity of the magnetic order in a way that resembles 
experiments.  From the comparison, we deduce that the coupling to the 
symmetry breaking field is small, $\varepsilon \approx 18\,{\rm mK }$,
compared to $T_{c+} \approx 540 \,{\rm mK}$.

The domain sizes consistent with specific heat experiments, 
$ 10 \xi_\varepsilon - 20 \xi_\varepsilon$ ($\sim 30\xi_0 -60 \xi_0$), are much larger 
than the uniformly sized $1\xi_\varepsilon - 2 \xi_\varepsilon$ ($\sim 3 \xi_0 - 6 \xi_0$) 
domains attributed to neutron scattering experiments.  For the latter small domains, 
only a single phase transition would appear in these calculations.  This reflects 
the relative stiffness of the superconducting order parameter as compared to the 
condensation energy gained from orienting the superconducting order parameter in 
a direction favorable to the local AFM order parameter.

%
%% fig 8
%

\subsection{Model II - `Swiss Cheese'}

The superconducting state of Model I is very sensitive to the size of the AFM 
domains.  To explore the response of an E$_{2u}$ superconductor to different spatial 
AFM configurations, we consider a model that represents what might be viewed as
an opposite extreme from the abutting uniformly sized domains of Model I.
In Model II, nanoscale sized AFM domains permeate a 
large single (`infinite') AFM 
domain like the holes in swiss cheese.
This picture is motivated by the observation of {\it intrinsic} defects
(most likely dislocation lines or stacking faults) in high quality 
and high purity UPt$_3$ samples\cite{walko00,midgley93,demczyk93} and
by the observation that neutron scattering data cannot distinguish between 
the commonly accepted picture of abutting uniform domains with small moments
and small domains with significantly larger moments.
These {\it intrinsic} defects may act as nucleation centers for 
the random-field-like symmetry breaking field, `nanodomains,' 
and may provide a natural explanation for the linewidth broadening of the AFM 
Bragg peak in reciprocal space as seen in neutron diffraction 
measurements.\cite{neutrons}

\begin{figure}
\noindent
\begin{minipage}{\hsize}
\epsfxsize=48mm
\centerline{{\epsfbox{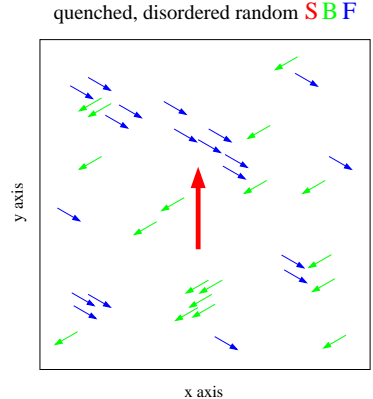}} }
\vspace*{0.1in}
\caption[]{The spatial orientation of the SBF for the
randomly dispersed nanodomains (small arrows) in the presence of a
uniform SBF in the background (\red{big arrow}) of Model II.
}
\label{m2:afm}
\end{minipage}
\end{figure}

\begin{figure}
\noindent
\begin{minipage}{\hsize}
\epsfxsize=53mm
\epsfysize=53mm
\centerline{\rotate[r]{\epsfbox{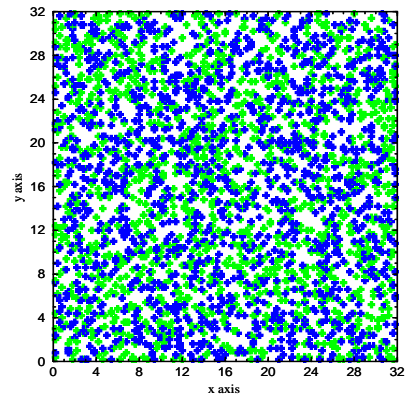}} }
\caption[]{A typical spatial distribution of cross-shaped nanodomains (nanodefects)
covering approximately 44\% of the $32\xi_\varepsilon \times 32\xi_\varepsilon$ 
numerical mesh in the presence of a uniform AFM background.
}
\label{nanodefects}
\end{minipage}
\end{figure}

\begin{figure}
\noindent
\begin{minipage}{\hsize}
\epsfysize=69mm
\centerline{\rotate[r]{\epsfbox{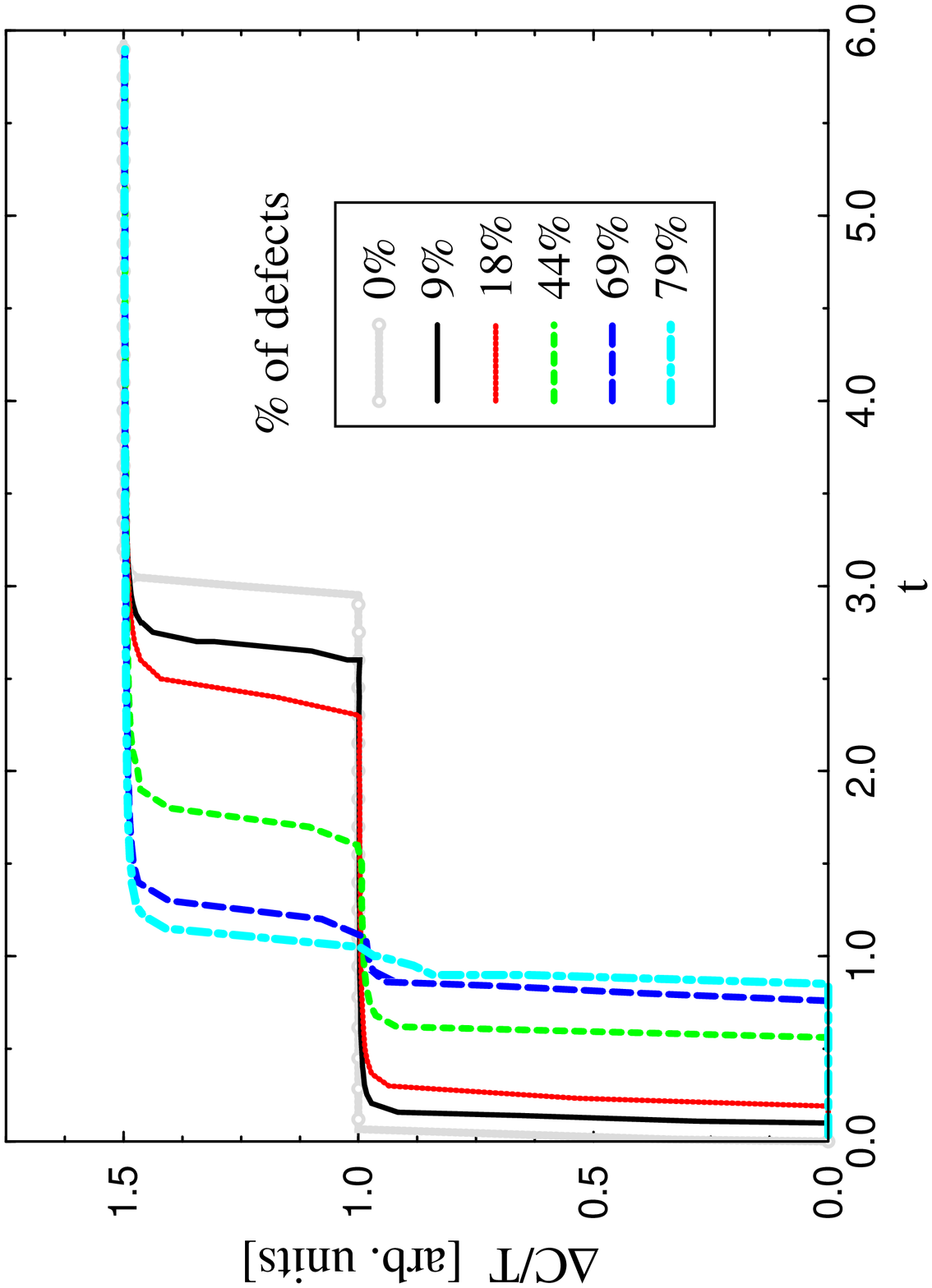}} }
\epsfysize=69mm
\centerline{\rotate[r]{\epsfbox{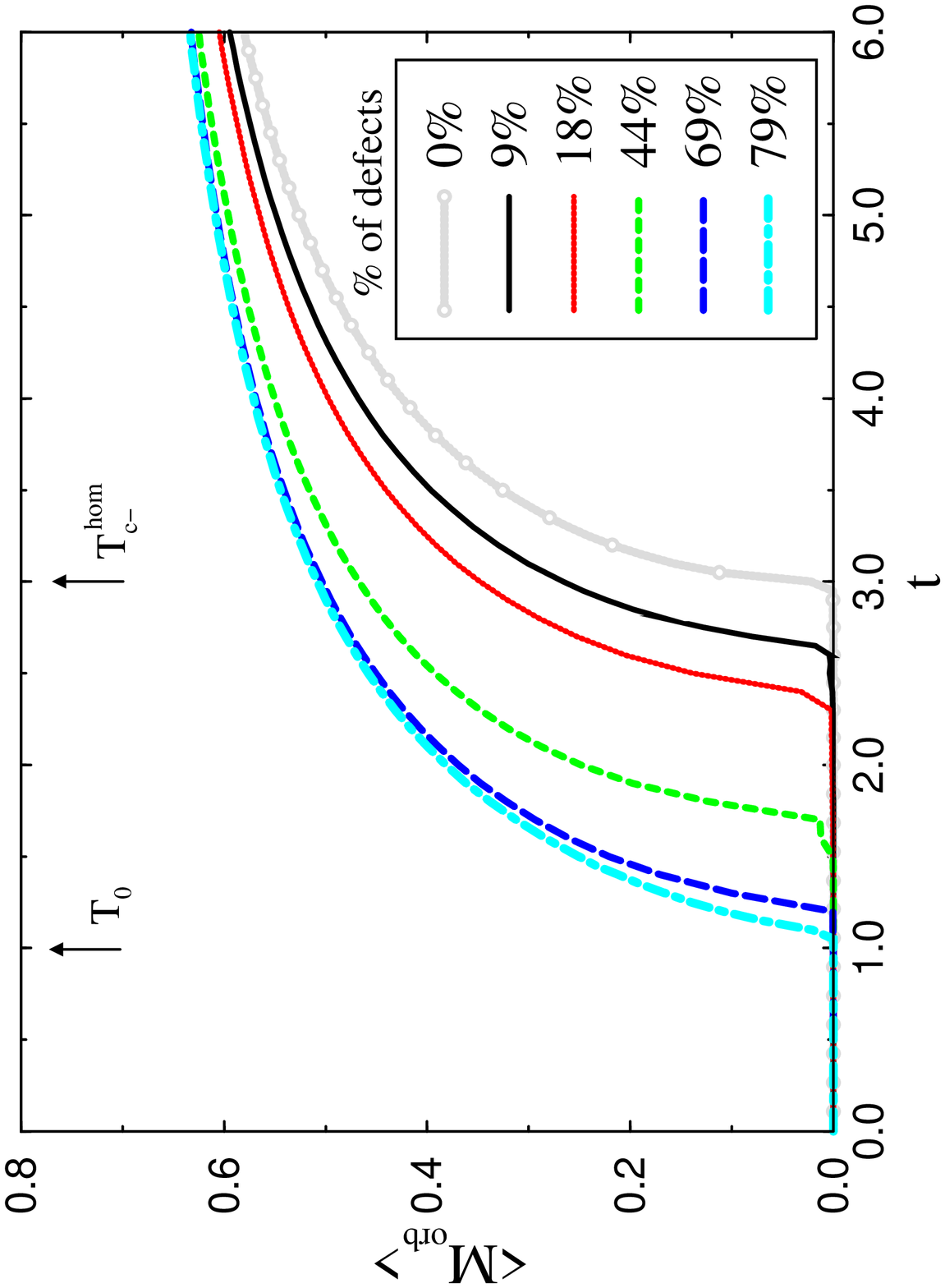}} }
\caption[]{Top:
The specific heat for Model II in units of 
$\triangle C_{+}^{\rm hom}/T_{c+}^{\rm hom}$
for the nanoscale defect model for various
defect densities on a 
$32 \, \xi_\varepsilon \times 32 \, \xi_\varepsilon$
lattice. 	
Bottom: The orbital magnetization for the same model 
and set of parameters as in the top panel.
}
\label{model_2_results}
\end{minipage}
\end{figure}

We used a standard pseudo-random number generator to uniformly distribute
the nanodomains on our computational mesh, and to assign a random 
orientation for the direction of ${\bf N}$ 
for each nanodomain relative to ${\bf N}$ of the `background' AFM
($\theta({\bf r})=\pm 120^\circ$).  On a mesh of $160\times 160$ points 
($32\xi_\varepsilon \times 32\xi_\varepsilon$)
approximately $3400$ nanodomains (nanodefects) are needed to cover all three 
orientations of the SBF equally.
An example configuration of the AFM order appears in Fig.~\ref{m2:afm}. 
In Fig.~\ref{nanodefects} we show a typical distribution of nanodefects
on a mesh of $32\xi_\varepsilon \times 32\xi_\varepsilon$ with a concentration
of defects that covers approximately 44\% of the mesh. Each 
nanoscale defect has a cross-shaped $5$-point layout on the mesh.  Interactions 
between adjacent defects or between clusters of defects were neglected.
Fig.~\ref{model_2_results} shows the specific heat as a function of temperature
obtained by heating from deep in the low temperature phase.  Two phase transitions 
are signaled by heat capacity jumps that remain sharp, consistent with second order 
phase transitions, even in the presence of a high density of nanoscale defects. 
Increasing nanodomain density does lead to a reduction in the splitting of
$T_c$ and only a single phase transition is observed for a nanodomain 
density larger than $\sim 75$\%.  The temperature evolution of the spatially 
averaged spontaneous magnetization shown in Fig.~\ref{model_2_results} indicates 
that the low temperature transition that separates two superconducting states 
is second order.  As expected, the rapid increase in the spontaneous magnetization 
is correlated with the appearance of a jump in the heat capacity, as shown
in the top panel of Fig.~\ref{model_2_results}.  Note that time-reversal 
symmetry is broken for the single transition that occurs at high nanodomain 
densities. 

%
%% fig 9
%

%
% fig 10
%

%
% fig 11 (a-b)
%

%
%% fig 12
%

\begin{figure}
\noindent
\begin{minipage}{\hsize}
\epsfxsize=58mm
\centerline{{\epsfbox{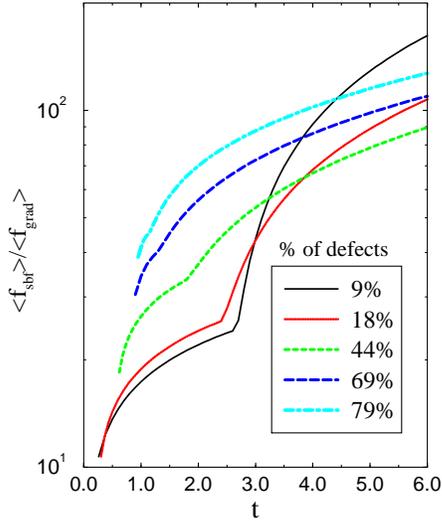}} }
\caption[]{The ratio of the contributions to the free energy from 
$\langle f_{\rm sbf}({\bf r}) \rangle_{\bf r}$
and $\langle f_{\rm grad}({\bf r}) \rangle_{\bf r}$. 
A large ratio reflects a stiff condensate; it is 
energetically less favorable for the superconducting 
order parameter to twist near the nanodomains.
Note that the second transition can be described by
$t_{-} \simeq \exp\big( [1 - x/x_{cr}]\ln 3 \big)$, 
where $x_{cr} \approx 88\%$  is the critical 
concentration of nanoscale defects,
where there is only one superconducting transition.
}
\label{sbf_grad_imps}
\end{minipage}
\end{figure}

In contrast to Model I, the contribution to the free energy from the 
symmetry breaking field dominates that from the gradient term as reflected 
in the large ratios 
$\langle f_{\rm sbf}({\bf r}) \rangle_{\bf r}/\langle f_{\rm grad}({\bf r}) \rangle_{\bf r}$
shown in Fig.~\ref{sbf_grad_imps}. The superconducting order 
parameter is on average aligned relative to the (background) AFM order 
except in a region $\sim \xi_{\varepsilon}$ around a nanodomain.

%
% fig 13
%

\begin{figure}
\noindent
\begin{minipage}{\hsize}
\epsfysize=72mm
\centerline{\rotate[r]{\epsfbox{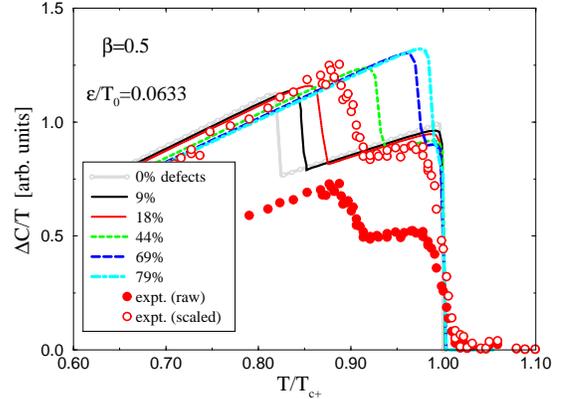}} }
\caption[]{Calculated specific heat as a function of 
temperature for Model II for various concentrations of nanodomains,
shown in comparison with experiment 
(filled circles) normalized by $C_N/T$.\cite{keizer99} 
For ease of comparison, the experimental data scaled by 
a numerical factor (open circles) is also presented.
The $T$-dependence of the GL result has been corrected 
according to Fig.~\ref{bcs-gl}.
}
\label{central2}
\end{minipage}
\end{figure}

In Fig.~\ref{central2} we compare our results of the specific heat with 
measurements on high-quality crystals.  Since the heat jumps are very
insensitive to the density of nanodomains, a wide range of $\varepsilon$
values and nanodomain concentrations are consistent with experiments.
In particular, an SBF coupling $\varepsilon/T_0 = 0.0633$ requires a 
concentration of roughly $36\%$ in order to account for the observed sharp
double transition in heat capacity measurements.

It is natural to expect that disorder will drive the lower temperature
phase transition from being second order to 
first order or possibly a glass transition. 
Our numerical heating and cooling cycles have shown that upon 
heating-up and crossing the low temperature phase transition, 
the entropy is always smooth and hence the transition is second order. 
However, when starting the cooling cycle above $T_{c-}$ we find a 
glasslike, frustrated and strongly disordered solution for the order 
parameter, which gives rise to a discontinuity in the entropy upon 
crossing $T_{c-}$. This discontinuity is consistent
with a first order transition. However, a comparison of the calculated
free energies shows that the latter is energetically less 
favorable than the solution with a smooth transition and signals 
that the glasslike solution is metastable. 
Assuming that the metastable glasslike solution is experimentally
observable when rapidly cooling down, our calculations give a
small latent heat 
$\ell = T_{c-} \triangle S = \mu \, T_{c-} \triangle C_{-}$,
where $\mu$ is a numerical factor of order $\mu \sim 0.01$. In other words,
the latent heat is a small fraction of the overall measured specific heat,
$Q \approx T_{c+} C_N \approx 200 \, {\rm mJ/mol}$, with
$\ell / Q  < 1\%$ or even less. In a carefully devised heat
capacity measurement this small latent heat should be observable
if indeed a glasslike phase transition occurs.

\section{Conclusions}

 We have explored the effect of spatially inhomogeneous antiferromagnetic order
coupled to the superconducting order parameter of an E$_{2u}$ superconductor in
two models representing limiting configurations of the AFM order parameter:   
(I) abutting AFM domains of uniform size equally distributed over the three 
 possible orientations of the AFM order parameter, and (II) small domains with
 dimensions of the order of the superconducting coherence length (nanodomains), 
 randomly dispersed through a single AFM domain.  Our numerical solutions of the 
 Ginzburg-Landau equations show that Model I is very sensitive to domain size.  
 Phase transitions are rapidly broadened and smeared as the domain size is decreased. 
 For domain sizes less than $\sim 2 \xi_{\varepsilon}$ only one transition is evident.  
 The results of our calculations in two-dimensions are in qualitative agreement with 
 those of the simple one-dimensional model of Garg. In contrast, Model II shows sharp 
 phase transitions for all nanodomain densities up to the point where the double
 phase transition gives way to a single phase transition.  Our calculations for 
 Model II show that low-lying metastable states affect the thermodynamic
 properties of the system as it is cooled, and suggest the possibility that the 
 lower transition may be weakly first order.  In contrast, our calculations for 
 Model I show no evidence of a first order transition.  
Although both models can yield sharp phase transitions like those
observed in the heat capacities of high-quality samples, neither of them
can account for the change in the heat capacity on annealing if the magnetic 
moments and domain sizes do not change as a result of annealing. 
The qualitative difference in the results for our two models
does caution that simplistic models involving domains {\em homogeneous} in size 
do not rule out the possibility of E$_{2u}$ superconductivity in UPt$_3$. 
The relative insensitivity of the low temperature time-reversal symmetry 
breaking phases to SIAFM in both models provides a natural explanation of how 
an E$_{2u}$ superconductor can provide a good description of the gap structure 
of UPt$_3$ at low temperature and therefore transport properties that are
in good agreement with experiment.  In the temperature region that includes the two
phase transitions, Model I appears to be too sensitive to SIAFM while 
Model II is perhaps not sensitive enough.  It is, thus, more 
likely that SIAFM is not arranged in abutting domains of uniform size, but
rather, there is a distribution of AFM domain sizes peaked around a particular 
size.  Further calculations are required to properly consider this possibility, 
which we examine in forthcoming work.

\acknowledgements

We thank J.A. Sauls for enlightening discussions early in this work and 
P. Kumar and H. R\"oder for many stimulating discussions.
M.J.G.  acknowledges the support of Los Alamos National Laboratory under 
the auspices of the Department of Energy and D.W.H. acknowledges the support 
of the Office of Naval Research.  This work was supported in part by a grant of
computer time from the DoD High Performance Computing and Modernization
Program on the Naval Research Laboratory's Origin 2000 and SUN Wildfire
computers.

\end{multicols}
 

\begin{thebibliography}{10}
\bibitem{neutrons} 
G. Aeppli, E. Bucher, C. Broholm, J.K. Kjems, J. Baumann, and
J. Hufnagl, Phys. Rev. Lett. {\bf 60}, 615 (1988).

\bibitem{isaacs95} 
E.D. Isaacs, P. Zschack, C.L. Broholm, C. Burns,  G. Aeppli, A.P. Ramirez, 
T.T.M. Palstra, R.W. Erwin, N. St\"{u}cheli, and E. Bucher, 
Phys. Rev. Lett. {\bf 75}, 1178 (1995).

\bibitem{hayden} 
S.M. Hayden, L. Taillefer, C. Vettier, and J. Flouquet,
Phys. Rev. B {\bf 46}, 8675 (1992);

\bibitem{adenwalla90} 
B.S. Adenwalla, S.W. Lin, Q.Z. Ran, Z. Zhao, J.B. Ketterson,
J.A. Sauls, L. Taillefer, D.G. Hinks, M. Levy, and B.K. Sarma,
Phys. Rev. Lett. {\bf 65}, 2298 (1990).

\bibitem{bruls90}
G. Bruls, D. Weber, B. Wolf, P. Thalmeier, and B. L\"{u}thi,
Phys. Rev. Lett. {\bf 65}, 2294 (1990).

\bibitem{uniax}
M. Boukhny, G.L. Bullock, B.S. Shivaram, and D.G. Hinks,
Phys. Rev. Lett. {\bf 73}, 1707 (1994);
Phys. Rev. B {\bf 50}, 8985 (1994);
D.S. Jin, S.A. Carter, B. Ellman, T.F. Rosenbaum, and D.G. Hinks, 
Phys. Rev. Lett. {\bf 68}, 1597 (1992).
             
\bibitem{sauls94} 
J.A. Sauls, Adv. Phys. {\bf 43},  113 (1994).

\bibitem{heffner96} 
R.H. Heffner and M.R. Norman, Comments Condens. Matter Phys. {\bf 17}, 
361 (1996).

\bibitem{machida89}
K. Machida, M. Ozaki, and T. Ohmi, J. Phys. Soc. Japan {\bf 58}, 4116 (1989).

\bibitem{chen93} 
D.-C. Chen and A. Garg, Phys. Rev. Lett. {\bf 70}, 689 (1993);
 A. Garg and D.-C. Chen, Phys. Rev. B {\bf 49}, 479 (1994).

\bibitem{cox95} 
R. Heid, Ya. B. Bazaliy, V. Martisovits, and D.L. Cox, 
Phys. Rev. Lett. {\bf 74}, 2571 (1995).

\bibitem{zhitomirsky96} 
M.E. Zhitomirsky and K. Ueda, Phys. Rev. B. {\bf 53}, 6591 (1996).

\bibitem{garg98} 
A. Garg, J. Phys.: Condens. Matter {\bf 10}, 4223 (1998).

\bibitem{keizer99} 
R.J. Keizer, A. de Visser, M.J. Graf, A.A. Menovsky,
and J.M. Franse, Phys. Rev. B {\bf 60}, 10527 (1999).

\bibitem{nosignatures} 
H. Tou, Y. Kitaoka, K. Asayama, N. Kimura, Y. Onuki, E. Yamamoto,
 and K. Maezawa, Phys. Rev. Lett. {\bf 77}, 1374 (1996);
P. Dalmas de R\'{e}otier and A. Yaouanc, Phys. Lett. A {\bf 205}, 239 (1995);
R.A. Fisher, B.F. Woodfield, S. Kim, N.E. Phillips, L. Taillefer,
 A.L. Giorgi, and J.L. Smith, Solid State Commun. {\bf 80}, 263 (1991); 
J.P. Vithayathil, D.E. Maclaughlin, E. Koster, D.L. Williams,
 and E. Bucher, Phys. Rev. B {\bf 44}, 4705 (1991);
M. Lee, G.F. Moores, Y.Q. Song, W.P. Halperin, W.W. Kim,
 and G.R. Stewart, Phys. Rev. B {\bf 48}, 7392 (1993).
 
\bibitem{muons2} 
A. Yaouanc, P.D. de~Reotier, F.N Gygax, A. Schenck, A. Amato,
 C. Baines, P.C.M. Gubbens, C.T. Kaiser, A. de~Visser, R.J. Keizer,
 and A. Huxley, Phys. Rev. Lett. {\bf 84}, 2702 (2000).

\bibitem{earlysh} 
A. Sulpice, P. Gandit, J. Chaussy, J. Flouquet, D. Jaccard, 
 P. Lejay, and J.L. Tholence, J. Low Temp. Phys. {\bf 62}, 39 (1986).

\bibitem{keizer99s} 
R.J. Keizer, A. de Visser, A.A. Menovsky, J.J.M. Franse, B. F{\aa}k, 
 and J.-M. Mignot, Phys. Rev. B {\bf 60}, 6668 (1999).

\bibitem{lussier96} 
B. Lussier, L. Taillefer, W.J.L. Buyers, T.E. Mason,
 and T. Peterson, Phys. Rev. B {\bf 54}, R6873 (1996).

\bibitem{vandijk98} 
N.H. van Dijk, B. F{\aa}k, L.P. Regnault,
 A. Huxley, and M-T. Fern\'{a}ndez-D\'{\i}az, 
 Phys. Rev. B {\bf 58}, 3186 (1998).

\bibitem{moreno} 
J. Moreno and J.A. Sauls,  e-print: cond-mat/0004253.

\bibitem{volovik88}
G.E. Volovik, J. Phys. C {\bf 21}, L221 (1988).

\bibitem{hess89} 
D.W. Hess, T.A. Tokuyasu, and J.A. Sauls, J. Phys.: 
 Condens. Matter {\bf 1}, 8135 (1989); Physica B {\bf 163}, 720 (1990).

\bibitem{tokuyasu90}
T.A. Tokuyasu, D.W. Hess, and J.A. Sauls, Phys. Rev. B {\bf 41}, 8891 (1990).

\bibitem{joynt90} 
R. Joynt, V.P Mineev, G.E. Volovik, and M.E. Zhitomirsky, 
 Phys. Rev. B {\bf 42}, 2014 (1990).

\bibitem{blount90}
E.I. Blount, C.M. Varma, and G. Aeppli, Phys. Rev. Lett. {\bf 64}, 3074 (1990)

\bibitem{norman92}
M.R. Norman, Physica C {\bf 194}, 203 (1992).

\bibitem{lukyanchuk93}
I. Luk'yanchuk and M.E. Zhitomirsky, Physica C {\bf 206}, 373 (1993).

\bibitem{park96} 
K.A. Park and R. Joynt, Phys. Rev. B {\bf 53}, 12346 (1996).

\bibitem{multi_machida} 
K. Machida and M. Ozaki, Phys. Rev. Lett. {\bf 66}, 3293 (1991);
T. Ohmi and K. Machida, Phys. Rev. Lett. {\bf 71}, 625 (1993);
 K. Machida, T. Nishira, and T. Ohmi, J. Phys. Soc. Jpn. {\bf 68},
 3364 (1999).


\bibitem{snote} 
A recent high-energy x-ray scattering experiment,
Ref.~\onlinecite{walko00}, claims that the original D$_{6h}$ symmetry 
classification for UP$_3$ is incorrect and that the correct crystal symmetry 
is trigonal.  The D$_{3h}$ symmetry group contains a single 2D 
E-representation.
The GL theory for a superconducting order parameter that transforms like this 
representation is formally identical to the E$_{2u}$ representation that we 
consider.  The calculations we present here are applicable without modification 
to an E$_{u}$ superconducting state of D$_{3h}$.

\bibitem{graf96} 
M.J. Graf, S.-K. Yip, and J.A. Sauls, J. Low Temp. Phys.
 {\bf 102}, 367 (1996); {\bf 106}, 727(E) (1997); {\bf 114}, 257 (1999).

\bibitem{graf99} 
M.J. Graf, S.-K. Yip, and J.A. Sauls, 
 Physica B {\bf 280}, 176 (2000).

\bibitem{choi91} 
C.H. Choi and J.A. Sauls, Phys. Rev. Lett. {\bf 66}, 
484 (1991); Phys. Rev. B {\bf 48}, 13\,684 (1993).

\bibitem{graf00}
M.J. Graf, S.-K. Yip, and J.A. Sauls, 
Phys. Rev. B {\bf 62}, 14393 (2000).
%e-print: cond-mat/0006180.


\bibitem{hess94} 
D.W. Hess, Physica B {\bf 194-196}, 1419 (1994).

\bibitem{sauls96} 
J.A. Sauls, Phys. Rev. B {\bf 53},  8543 (1996).

\bibitem{keller94} 
N. Keller, J.L. Tholence, A. Huxley, and J. Flouquet,
Phys. Rev. Lett. {\bf 73}, 2364 (1994); {\bf 74}, 2148 (E) (1995).

\bibitem{mineev91} 
V.P. Mineev, Physica B {\bf 171}, 138 (1991).

\bibitem{grafnext} M.J. Graf and D.W. Hess (unpublished).

\bibitem{aeppli94} 
G. Aeppli and C. Broholm in {\it Handbook on the
Physics and Chemistry of Rare Earths}, eds. K.A. Gschneider 
{\it et al.} (Elsevier Science B. V., 1994), Vol. 19, 123.

\bibitem{footnote_AFM}
It is worth noting that Eq. (\ref{f_sbf}) is by design traceless
and does not include a possible coupling term proportional to 
$|{\veta}|^2|{\bf N}|^2$.  This term is implicitly absorbed into the 
definition of $T_0$, as has been done in Ref.~\onlinecite{hess89}.
An explicit accounting of this term is not relevant to the central focus 
of this work and does not affect its conclusions. Since this term is
important in determining the temperature dependence of $|{\bf N}|$ below the
superconducting phase transition
[see for example, the superconducting glass model of 
B. Kishore and P. Singh, Physica C {\bf 215}, 59 (1993)],
such a term must be explicitly included in any self-consistent model 
of coupled E-representation superconductivity and AFM order.  Such a 
program was carried out for the single-domain E-rep models of 
Ref.~\onlinecite{park96} to make a quantitative analysis of the H-P-T phase 
diagram. Our work suggests that a quantitative E-rep
model of the H-P-T phase diagram should take the spatial dependence of the 
AFM order into account; one cannot assume that the equations of a 
single-domain model with `renormalized couplings' will be valid. 
 
\bibitem{sauls94a}
J.A. Sauls, J. Low Temp. Phys. {\bf 95}, 153 (1994).

\bibitem{spec_heat} 
J.P. Brison, N. Keller, P. Lejay, J.L. Tholence, A. Huxley,
N. Bernhoeft, A.I. Buzdin, B. Fak, J. Flouquet, and
L. Schmidt, J. Low Temp. Phys. {\bf 95}, 145 (1994).

\bibitem{kleiman92}
R.N. Kleiman, C. Broholm, G. Aeppli, E. Bucher, N. St\"ucheli, D.J. Bishop,
K.N. Clausen, K. Mortensen, J.S. Pedersen, and B. Howard,
Phys. Rev. Lett. {\bf 69}, 3120 (1992).

\bibitem{gcomp} 
Note that the specific heat curves in Ref.~\onlinecite{garg98} are 
labeled by a parameter that corresponds to the square of the domain size in our units. 

\bibitem{smallcomp} 
Small computational meshes may artificially suppress a
transition to a glass phase.

\bibitem{glassnote} 
This phase is apparently not a glass, 
because $\langle \eta_i \rangle_{\bf r} \neq 0$ according to the 
Edwards-Anderson definition of a superconducting glass order parameter.

\bibitem{walko00}
D.A. Walko, J.-I. Hong. T.V.C. Rao, Z. Wawrzak, D.N. Seidman,
W.P. Halperin, and M.J. Bedzyk, preprint 2000.

\bibitem{midgley93} 
P.A. Midgley, S.M. Hayden, L. Taillefer, B. Bogenberger, 
and H. von L\"{o}hneysen, Phys. Rev. Lett. {\bf 70}, 678 (1993).

\bibitem{demczyk93}
B.G. Demczyk, M.C. Aronson, B.R. Coles, and J.L. Smith,
Philos. Mag. Lett. {\bf 67}, 85 (1993);

\end{thebibliography}
\end{document}